\documentclass[lettersize,journal]{IEEEtran}
\usepackage{amsmath,amsfonts}
\usepackage{algorithmic}
\usepackage{array}
\usepackage[caption=false,font=normalsize,labelfont=sf,textfont=sf]{subfig}
\usepackage{textcomp}
\usepackage{stfloats}
\usepackage{url}
\usepackage{verbatim}
\usepackage{graphicx}
\usepackage{amssymb}
\def\BibTeX{{\rm B\kern-.05em{\sc i\kern-.025em b}\kern-.08em
    T\kern-.1667em\lower.7ex\hbox{E}\kern-.125emX}}
\usepackage{balance}

\usepackage{cite}

\usepackage{algorithm}

\begin{document}
\title{
Active RIS-Aided Massive MIMO  
With
Imperfect CSI and
Phase Noise 

	 }
\author{ Zhangjie Peng, Jianchen Zhu, Cunhua Pan,~\IEEEmembership{Senior Member,~IEEE},
	Zaichen Zhang,~\IEEEmembership{Senior Member,~IEEE},\\
	Daniel Benevides da Costa,~\IEEEmembership{Senior Member,~IEEE},
	Maged Elkashlan,~\IEEEmembership{Senior Member,~IEEE},\\
	and	
	George K. Karagiannidis,~\IEEEmembership{Fellow,~IEEE}

\thanks{
Z. Peng and J. Zhu
are with the College of Information, Mechanical and Electrical Engineering,
Shanghai Normal University, Shanghai 200234, China (e-mail: pengzhangjie@shnu.edu.cn; 1000526997@smail.shnu.edu.cn).
}
\thanks{
C. Pan and Z. Zhang
are with the National Mobile Communications
Research Laboratory, Southeast University, Nanjing 210096, China (e-mail:
cpan@seu.edu.cn; zczhang@seu.edu.cn).
}
\thanks{
D. B. da Costa
is with the Department of Electrical Engineering, King Fahd
University of Petroleum 
$\&$ Minerals, Dhahran 31261, Saudi Arabia
(e-mail: danielbcosta@ieee.org).
}
\thanks{
	M. Elkashlan
	 is with the School of Electronic Engineering and Computer
	Science, Queen Mary University of London, E1 4NS London, U.K. (e-mail:
	maged.elkashlan@qmul.ac.uk).
}
\thanks{
	G. K. Karagiannidis 
	is with the Department of Electrical and Computer Engineering, Aristotle University of Thessaloniki, 54124 Thessaloniki, Greece, 
	and also
	with the Artificial Intelligence $\&$ Cyber Systems Research Center, Lebanese
	American University (LAU), Lebanon 
	(e-mail:geokarag@auth.gr).
}
}

\markboth{}%
{How to Use the IEEEtran \LaTeX \ Templates}

\makeatletter
\newcommand{\rmnum}[1]{\romannumeral #1}
\newcommand{\Rmnum}[1]{\expandafter\@slowromancap\romannumeral #1@}
\makeatother

\makeatletter
\renewcommand{\maketag@@@}[1]{\hbox{\m@th\normalsize\normalfont#1}}%
\makeatother

\newtheorem{remark}{Remark}
\newtheorem{theorem}{Theorem}
\newtheorem{corollary}{Corollary}
\newtheorem{lemma}{Lemma}
\maketitle

\setlength{\abovedisplayskip}{1pt}
\setlength{\belowdisplayskip}{1pt}
\setlength{\abovedisplayshortskip}{6.5pt}
\setlength{\belowdisplayshortskip}{6.5pt}

\begin{abstract}
Active reconfigurable intelligent surface (RIS)
has attracted significant attention as a recently proposed RIS architecture.
Owing to its capability to amplify the incident signals, 
active RIS can mitigate the multiplicative fading effect inherent in the passive RIS-aided system.
%
In this paper, we consider an active RIS-aided uplink multi-user massive multiple-input multiple-output (MIMO) system 
in the presence of
phase noise at the active RIS. 
Specifically, we employ a two-timescale scheme, 
where the beamforming at the base station (BS) is adjusted based on the
instantaneous aggregated
channel state information (CSI) and 
the statistical CSI serves as the basis for designing the phase shifts
at the active RIS, so that the feedback overhead  and computational complexity can be significantly  reduced. 
%
The aggregated channel composed of the cascaded and direct channels is estimated by utilizing 
the linear minimum mean square error (LMMSE) technique.
Based on the estimated channel, we derive the analytical closed-form expression of a lower bound of the
achievable rate.
The power scaling laws in the active RIS-aided system are investigated based on the theoretical expressions.
When the transmit power of each user is scaled down by the number of BS antennas $M$ or reflecting elements $N$,
we find that the thermal noise will cause the lower bound of the achievable rate to approach zero,
as the number of $M$ or $N$ increases to infinity. 
Moreover, an optimization approach based on genetic algorithms (GA) is introduced to tackle the phase shift optimization problem. 
Numerical results reveal that the active RIS can greatly enhance the  performance of the
considered system
 under various settings.

\end{abstract}


\section{Introduction}
\IEEEPARstart{R}{ecently}, 
the reconfigurable intelligent
surface (RIS) 
sparks considerable attention due to its deployment for attaining superior system performance \cite{ref1,ref2,ref3}.
Specifically, the RIS comprises numerous cost-effective and passive reconfigurable elements. These passive elements can intelligently adjust the phase shifts of the impinging waves by a  controller.
In addition, the RIS operates without the need for 
radio frequency (RF) chains
and
digital signal processing circuits, 
allowing for 
a thin construction that  facilitates easy installation
on building facades and indoor spaces\cite{ref4}.
%
%

On the other side,
massive multiple-input multiple-output (MIMO) constitutes a technology within 
fifth generation (5G) 
communication standards,
which allows to serve multiple users simultaneously by spatial multiplexing \cite{sanguinetti2020toward}.
With a considerable number of
antennas deployed at the base station (BS), the massive MIMO system can achieve performance enhancement
\cite{chen2021analysis}.
Based on the aforementioned benefits of RISs, the combination of massive 
MIMO systems and RISs presents a compelling solution for future communication needs.
Hence, the research of RIS-aided massive MIMO communication systems has drawn significant attention from academia and industry.
In \cite{nguyen2022uav}, unmanned aerial
vehicles are integrated into the RIS-aided massive MIMO system, extending the network coverage. 
The authors of \cite{wang2021joint} demonstrated that deploying the RIS into the massive MIMO system can achieve great improvements in spectral efficiency by jointly optimizing the active and passive beamforming.
Moreover, an alternating
optimization framework 
was proposed in \cite{he2022reconfigurable}, 
maximizing channel capacity through
joint  passive beamforming and antenna selection
in RIS-aided massive MIMO 
systems.

%

In order to leverage the advantages offered by these additional antennas, 
the channel state information (CSI) is necessary for the operation of
communication systems. 
However,  previous works only considered  designing the phase shifts of the RIS based on
instantaneous CSI, and such scheme brings two challenges.
Specifically, one is that the overhead of obtaining instantaneous CSI
is 
prohibitively high
in  RIS-aided massive MIMO systems
given the large amount of reflecting elements \cite{bjornson2020intelligent}.
The other concerns the calculation of 
the optimal beamforming coefficients of the RIS, 
which need to be transmitted to the RIS controller via a dedicated feedback
link
\cite{ref3}. 
Due to the dynamic and evolving nature of wireless environments, 
 the information feedback  and the beamforming calculation require to be  frequently executed, which 
leads to a high computational complexity and energy consumption in the instantaneous CSI-based scheme.
Given the aforementioned drawbacks, 
a novel approach based on a two-timescale scheme was proposed in
\cite{ref15,ref16,ref17}.
In particular, the instantaneous CSI of the aggregated channel
is independent of the number of RIS elements and it can be
used in designing the transmit beamforming at the BS.
Besides, the configuration of the RIS phase shifts is dependent upon the long-term statistical CSI which varies slowly and does not require frequent updates in each channel coherence internal. 
As a result,
a reduction in computational complexity and estimation overhead can be obtained
 by using statistical CSI
\cite{ref13,2022zhikangdapower,zhi2022two-timescale}.


However, implementing the RIS can yield limited performance enhancements due to the “multiplicative fading" or “double fading" effect. 
As we all know,
the signals of the transmitter 
 need to pass through the cascaded channel, i.e., the 
 transmitter-RIS-receiver link in RIS-aided systems. 
Herein, the equivalent path-loss of the cascaded link is treated as a product
of the path-loss of the links at both ends of the RIS.
This leads to the path-loss of the cascaded link
being comparatively larger than that of the direct link\cite{ref5}. 
Therefore, the received signals are vulnerable to the multiplicative fading effect.

To overcome the limitations on the potential 
of traditional 
passive RISs, the authors
of \cite{ref6}  introduced a novel RIS architecture, named active RIS, 
where an active reflection-type amplifier is integrated into each RIS element.
While retaining the ability to adjust phase shifts like passive RISs, the significant capability of active RISs lies in amplifying the incident signals.
This new RIS architecture  offers a promising solution to counteract multiplicative fading effect.
Consequently, 
there has been a surge of research contributions in the literature.
The authors of \cite{ref9} studied the secure transmission in an active RIS-aided multi-antenna
system.
In \cite{zhi2022activeris},
a comparison between the active RIS-aided 
single-input single-output (SISO) system and the passive counterpart was conducted
 under the same power budget.
The authors of \cite{ma2022active} investigated the energy efficiency of  active RIS-aided multi-user multiple-input single-output (MU-MISO) systems. 
%


However, an unavoidable problem in active RIS-aided systems has been overlooked by all of the aforementioned works:
the
phase noise arising from imperfect configuration of the phase shifts at the RIS.
Hence, it is imperative to account for the phase noise present
at the RIS
in practical RIS-aided systems.
In \cite{2023liperformance}, 
the performance degradation induced by the phase noise 
at the active RIS
was
demonstrated
in the active RIS-aided SISO system.
The authors of \cite{liuxue2022_D2D}  derived the approximate closed-form expression of the ergodic sum rate of  an active RIS-aided multi-pair device-to-device system with phase noise.
However, research on integrating the active RIS into the massive MIMO
system while taking the phase noise at the RIS into account remains largely unexplored.

Against the aforementioned backgrounds,
we focus on an active RIS-aided uplink massive MIMO wireless communication system in the presence of imperfect CSI and phase noise at the active RIS.
In contrast to prior research
\cite{ref9,zhi2022activeris,ma2022active,2023liperformance,liuxue2022_D2D}, we 
take into account 
a substantial number of antennas deployed at the BS, which is a more complex scenario to adapt for the future communications. 
Besides, we adopt the two time-scale scheme in the considered system to reduce the 
channel estimation overhead, feedback overhead and the computational complexity considering the 
significant quantity
of  BS antennas and reflecting elements.
The main contributions can be 
outlined as follows:
\begin{itemize}
	\item[$\bullet$] 
	We investigate  an active RIS-aided uplink massive MIMO wireless communication system considering
	the phase noise at the active RIS and
	the direct links between the users and the BS. 
	
	\item[$\bullet$]
	With imperfect CSI,
	we employ a 
	linear minimum mean
	square error 
    (LMMSE) 
	algorithm for estimating the aggregated channel including the cascaded and direct links. Based on the estimated aggregated channel, the maximum ratio combining (MRC) technique 
	 is  adopted
	to the received signals at the BS.
	\item[$\bullet$] We obtain the closed-form expressions dependent on statistical CSI for the lower bound of the achievable rate in the  active RIS-aided massive MIMO system.
	Then, the power scaling laws  of the considered system in different scenarios
	are analyzed to obtain meaningful conclusions. 
    \item[$\bullet$]To guarantee fairness among multiple users, we
    address the problem of maximizing the minimum user rate
    by utilizing a genetic algorithm (GA). 
    Simulation results 
    reveal that the active RIS can significantly enhance the system performance. 
\end{itemize}

The remainder of this paper is structured as follows.  Section \uppercase\expandafter{\romannumeral 2} describes
the system model of an active RIS-aided massive MIMO system.
In Section \uppercase\expandafter{\romannumeral 3}, 
we perform the  derivation and analysis of  channel estimation based on LMMSE. The expression of the UatF bound of the achievable rate and analysis of the corresponding power scaling laws are included in Section \uppercase\expandafter{\romannumeral 4}.
In Section \uppercase\expandafter{\romannumeral 5}, the optimization problem for maximizing the minimum user rate is resolved based on a GA. 
Section \uppercase\expandafter{\romannumeral 6} provides the numerical results. Section 
\uppercase\expandafter{\romannumeral 7} presents a brief conclusion.

 {\it {Notations:}} Vectors and matrices are represented by lowercase and uppercase symbols,
  respectively. The  inverse, transpose and conjugate transpose of a matrix are represented by
  $\left( \cdot \right)^{-1}$, $\left( \cdot \right)^T$ and $\left( \cdot \right)^H$, respectively. Furthermore, 
  we use    
$\mathbb{E}\left\{ \cdot \right\}$, Tr$\left\{ \cdot \right\} $ and Re$\left\{ \cdot \right\}$ to represent the expectation, trace and the real part of a complex
number, respectively.
Also, 
diag (${\rm {\bf X}}$) denotes a diagonal matrix composed of 
elements from the diagonal of 
${\rm {\bf X}}$.
 $\left| \cdot  \right|$ and  $\left \| \cdot  \right \|$denote
 the absolute value of a complex
number and the norm of a vector, respectively.
${\mathbf I}_M$ denotes a $M \times  M$ identity matrix.
A vector 
denoted by $ \mathcal{C}\mathcal{N}\left( 0,\mathbf {\Sigma} \right)$
follows a circularly symmetric complex Gaussian distribution, with zero mean and a covariance matrix $ \mathbf {\Sigma}$.

\begin{figure}[!t]
	\centering
	\includegraphics[width=3.5in]{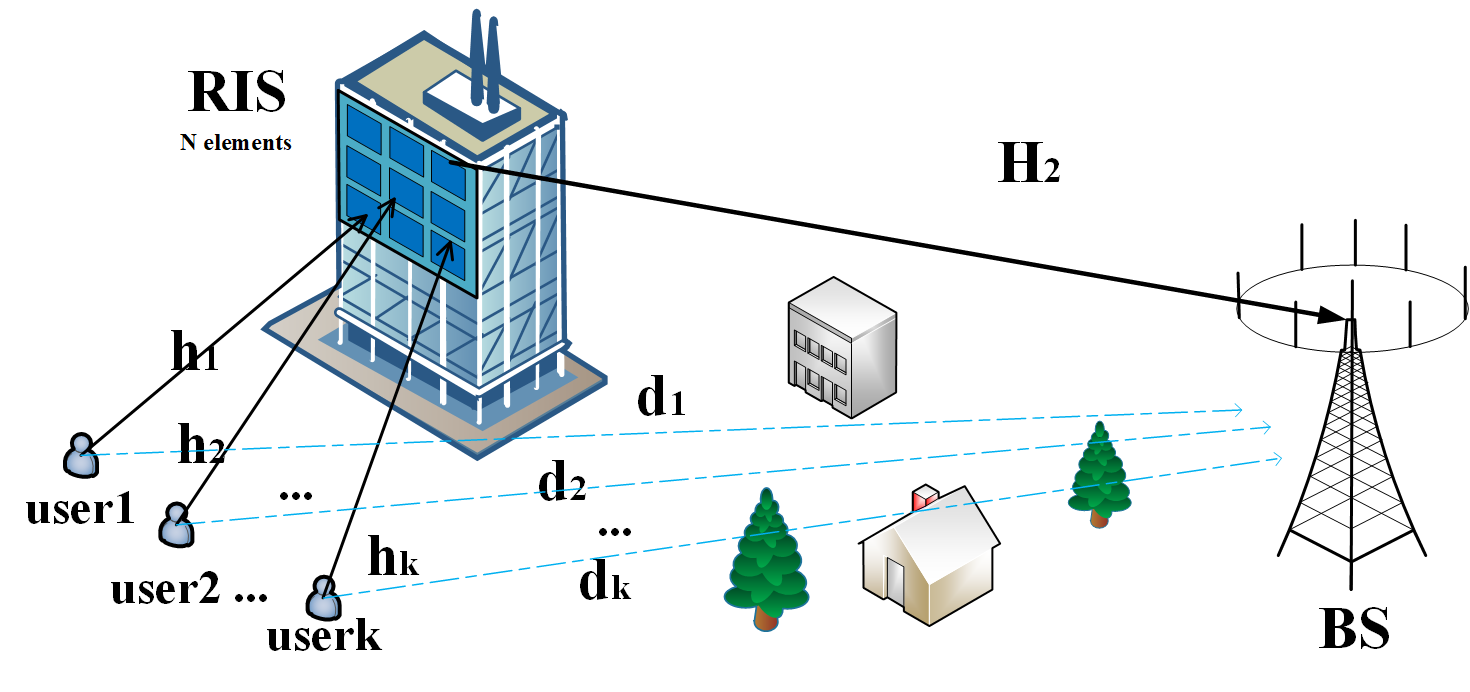}
	\caption{An active RIS-aided uplink communication system.}
	\label{fig_1}
\end{figure}
\section{System Model }
Our focus lies on the uplink transmission of an active RIS-aided massive MIMO system which is
illustrated in Fig. 1. Specifically, there are $K$ single-antenna users near the active RIS communicating with a BS. In this system, the active RIS comprises $N$ reflecting elements and the BS is equipped with $M$ antennas. 
 Moreover, there are also direct links from the users to the BS.


%

As depicted in Fig. 1,
the direct link from user $k$
to the BS, the channel from user $k$ to the active RIS, and the channel from the active RIS to the BS are denoted by ${\mathbf{d}_{ k}} \in {{\mathbb C}^{{M} \times {1}}}$, ${\mathbf{h}_{ k}} \in {{\mathbb C}^{{N} \times {1}}}$ and ${\mathbf{H}_{\rm 2}} \in {{\mathbb C}^{{M} \times {N}}}$, respectively. 
Also, the phase shift matrix of
the active RIS is represented by 
$\mathbf{\Phi }\!=\!\text{diag}\left\{ e^{j\theta _1},
...,e^{j\theta _N} \right\}
\in   {{\mathbb C}^{{N} \times {N}}}
$, 
where $\theta _n$ denotes the phase shift of the $n$-th reflection unit within the interval $\left[ 0,2\pi \right)$. 
The reflection coefficient matrix of the active RIS is given by
${\rm{\mathbf A }}=\text{diag}\left\{ \eta_{ 1},...,{\eta_{ N} }\right\}
\in   {{\mathbb C}^{{N} \times {N}}}$, where 
$\eta_{ n}$ denotes the amplification factor of the $n$-th RIS element,
and its value  can surpass one
due to the amplifiers of the active RIS.
For the simplification of subsequent research, we assume that 
$\eta_{ n}=\eta$.

 

Given the physical positions of the RIS and the BS in Fig. 1, we employ the Rician fading model to characterize the user-RIS channel and the RIS-BS channel respectively as follows
\begin{align}\label{channel_Users-RIS}
	&\mathbf{h}_{k}=\sqrt{\frac{\alpha _{k}}{\varepsilon _{k}+1}} \left (\sqrt {\varepsilon _{k}} \,\mathbf{\overline{h}}_{k}+{\tilde{\mathbf h}}_{k} \right),{k}\in\mathcal{K},\\
	&\mathbf{H}_{2}=\sqrt{\frac{\beta}{\delta+1}} \left (\sqrt {\delta}\, {\mathbf{\overline{H}}_2}+{{\tilde{\mathbf H}}}_{2} \right),
\end{align}
where 
$\alpha _{k}$ and $\beta$ are the path-loss coefficients that vary with distance, whereas $\varepsilon _{k}$ and
$\delta$ stand for the Rician factors of corresponding paths. 
Meanwhile, the set of users is denoted by $\mathcal{K} \!=\! \left\{{1, 2, . . . , K}\right\}$.
Moreover, $\mathbf{\overline{h}}_{k}$ and $\mathbf{\overline{H}}_{2}$ signify the 
line-of-sight (LoS) components, whereas ${\tilde{\mathbf h}}_{k}$ and ${\tilde{\mathbf H}}_{2}$ represent the 
non-LoS
(NLoS) components. Specifically, the elements of ${\tilde{\mathbf h}}_{k}$ and ${\tilde{\mathbf H}}_{2}$  follow independent and identically distributed (i.i.d.) complex Gaussian distributions, each with a mean of zero and a variance of one. 
In this paper, the BS and the RIS  both adopt  the uniform square planar arrays (USPA). 
Therefore, we can  express
${\overline{\mathbf h}}_{k}$ and $\mathbf{\overline{H}}_{2}$  
 as follows 
\begin{align}\label{channel_User-RIS}
	&{\overline{\mathbf h}}_{k}=\mathbf{a}_N\left( \varphi _{k,{r}}^{a},\varphi _{k,{r}}^{e} \right),{k}\in\mathcal{K},\\
	\vspace{-0.4cm}
	&\mathbf{\overline{H}}_2=\mathbf{a}_M\left( \phi  _{{r}}^{a},\phi   _{{r}}^{e} \right) \mathbf{a}_{N}^{\mathrm{H}}\left( \varphi  _{{t}}^{a},\varphi_{{t}}^{e} \right),
\end{align}
where $\varphi _{k,{r}}^{a}(\varphi _{k,{r}}^{e})$ is the azimuth (elevation) angle of arrival (AoA) at the RIS from user $k$, 
$\varphi  _{{t}}^{a}(\varphi_{{t}}^{e})$
 is the azimuth (elevation) angle of departure (AoD) reflected towards the BS by the RIS. $\phi_{{r}}^{a}(\phi_{{r}}^{e})$ is the azimuth
 (elevation) AoA 
 from the RIS to the BS,
 respectively.
Besides, the $i$-th entry of the array response vector $\mathbf{a}_X(\vartheta^a,\vartheta^e) \in {{\mathbb C}^{{X} \times {1}}}$, $X \in \left\{M,N\right\}$, can be described as
\begin{align}\label{array_response_vector}
	[{{{\mathbf{a} }}_X}\left( {{\vartheta ^a},{\vartheta ^e}} \right)]_i = &{\rm{exp}}
	\Big \{
		j2\pi \frac{d}{\lambda }\left( {\left\lfloor {(x - 1)\sqrt {\mathrm{X}} } \right\rfloor} \sin {\vartheta ^e}\sin {\vartheta ^a} \notag
		\right.\\& \left.
		+\left( (x - 1)\rm{mod}\sqrt X \right)\cos {\vartheta ^e}\right)
	\Big \},
\end{align}
where $d$ stands for the antenna or element spacing, with $\lambda$ representing the wavelength. Furthermore, we use $\mathbf{a}_M$ and $\mathbf{a}_N$ to 
simplify the representation of
$\mathbf{a}_M\left( \phi  _{{r}}^{a},\phi   _{{r}}^{e} \right)$ and 
$\mathbf{a}_{N}\left( \varphi  _{{t}}^{a},\varphi_{{t}}^{e} \right)$
 respectively.

Considering the presence of various obstructions like trees and buildings between the users and the BS, the direct link between user $k$ and the BS experiences Rayleigh fading,  which is depicted as
\begin{align}
	{{\bf{d}}_k} = \sqrt {\gamma _k}{{\tilde{\bf d} }_k}, {k}\in\mathcal{K},
\end{align}
where $\gamma _k$ is the large-scale path-loss factor. 
The entries of ${{\tilde{\bf d} }_k}$ are i.i.d. complex Gaussian random variables, i.e.,  ${{\tilde{\bf d} }_k}\sim \mathcal{C}\mathcal{N}\left( 0,\mathbf {I}_M \right)$.


Herein, we consider the imperfect hardware at the active RIS.
Specifically, given the inherent limitations in configuring the reflection phases of RIS with limited precision, they can be characterized as phase noise \cite{2020badiucommunication}.
Therefore, the hardware impairment at the active RIS is described in a random diagonal matrix comprised of $N$ random phase noise \cite{papazafeiropoulos2021intelligent}, and we have
\begin{align}
	\mathbf{\Theta}=\text{diag}\left\{ e^{j\tilde{\theta }_1},e^{j\tilde{\theta }_2},...,e^{j\tilde{\theta }_N} \right\}\in {{\mathbb C}^{{N} \times {N}}}
	\label{Theta}.
\end{align}
In (\ref*{Theta}), $\tilde{\theta }_n$ follows a Von Mises distribution with a mean of zero whose 
probability density function (PDF) is
$M_v( \tilde{\theta})=\frac{e^{vcos(\tilde{\theta})}}{2\pi I_0{(v)}}$, 
with the concentration parameter $v$. 
The characteristic function of $\tilde{\theta}$
can be obtained as 
$\mathrm{E}\{ e^{j\tilde{\theta }}\}\!=\!\frac{I_1{(v)}}{I_0{(v)}}\triangleq\rho$,
$\mathrm{E}\{ e^{j2\tilde{\theta }}\}\!=\!\frac{I_2{(v)}}{I_0{(v)}}\triangleq l$,
where $I_n{(v)}$ represents the modified Bessel function of the first
kind and  $n$ represents its order.

 Considering the phase noise at the RIS in transmission, we can obtain the cascaded user-RIS-BS channel as
$\textbf{G}=
{\left[\textbf{g}_{1},\textbf{g}_{2}...,{\mathbf{g}}_{K}\right]}=\mathbf{H}_{\rm 2}\mathbf{A \Phi }
{\mathbf{\Theta}}
{\mathbf{H}}_{\rm 1}\!\in\! {{\mathbb C}^{{M} \times {K}}}$ with
${{\textbf{g}}_{ k}}\!\!=\!\mathbf{H}_{\rm 2}\mathbf{A\Phi }{\mathbf{\Theta}}{\mathbf{h}}_{ k}$, where $\mathbf{H}_{\rm 1}\!=\!{\left[\mathbf{h}_{\rm 1},\mathbf{h}_{\rm 2}...,\mathbf{h}_{ K}   \right]}$.
As a result, the aggregated channel from user $k$ to the BS can be expressed as 
\begin{align}
	{{\bf{q}}_k} &=\,  {{\bf{g}}_k} + {{\bf{d}}_{{k}}} = {{\bf{H}}_{{2}}}{\bf{ A\Phi\Theta }}{{\bf{h}}_{{k}}} + {{\bf{d}}_{{k}}}\notag 
	\\
	&= 
	\underbrace 
	{ \sqrt {{c_k}\delta {\varepsilon _k}} {{{\bf{\overline H}}}_2}{\bf{A\Phi\Theta }}{{{ {\overline {\mathbf h}}}}_{{k}}}}
	_{{\bf{q}}_k^1} 
+ 
	\underbrace 
{ \sqrt {{c_k}\delta } {{{\bf{\overline H}}}_2}{\mathbf{A\Phi\Theta } 
		\tilde {\mathbf h}_k}^{\rm{}}}
_{{\bf{q}}_k^2}
\notag  
	\\
	&	+  
	\underbrace 
	{\sqrt {{c_k}{\varepsilon _k}} {{{\tilde{ \mathbf H}}}_2}{\bf{A\Phi\Theta }}{{{\bf{\overline h}}}_{{k}}}}
	_{{\bf{q}}_k^3} 
	+  
	\underbrace 
	{\sqrt {{c_k}} {{{\tilde{ \mathbf H}}}_2}{\bf{A\Phi\Theta }}{{{\tilde{ \mathbf h}}}_k}}
	_{{\bf{q}}_k^4} 
	+ \sqrt {{\gamma _k}} {{\tilde{\bf d} }_k}\label{q_k},
\end{align}
where ${c_k}\triangleq{\frac{\beta\alpha _{k}}{(\delta+1)(\varepsilon _{k}+1)}}$.
We define the first four terms of (8) as
 ${{\bf{q}}_k^1} \sim {{\bf{q}}_k^4}$, then
 	${{\bf{q}}_k}	=\sum\limits_{w = 1}^4 {\mathbf{q}_k^w} + \mathbf{d}_k
 	\triangleq  {\underline{\bf{{ q}}}_k} + \mathbf{d}_k$.
 	Therefore, the aggregated channel matrix from  users to the BS can be signified by $\mathbf{Q}\!=\!\textbf{G}+\textbf{D}\!=\!{\left[\mathbf{q}_{\rm 1},\mathbf{q}_{\rm 2}...,\mathbf{q}_{ K}   \right]}\in {{\mathbb C}^{{M} \times {K}}}$, where 
 	$\mathbf{D}\!=\!{\left[\mathbf{d}_{ 1},\mathbf{d}_{ 2}...,\mathbf{d}_{K}   \right]}$.
Hence, the received signal vector at the BS is formulated as 
\begin{align}
	{\mathbf{y}}=\sqrt p{\mathbf{Qx}}
	=\sqrt p \sum\limits_{k = 1}^K\mathbf{q}_kx_k + \mathbf{H}_2 \mathbf{A\Phi\Theta}	
	 \boldsymbol{\nu} +  \mathbf{n},
\end{align}
where the transmit power of each user  is denoted by $p$,
$\mathbf{x}={\left[x_{\rm 1},x_{\rm 2}...,x_{K}  \right]}^T\!$ contains the transmit signals of all users, $\boldsymbol{\nu}\sim \mathcal{C}\mathcal{N}\left( 0,\sigma _{e}^{2}\textbf{I}_N \right)$ represents the thermal
noise which
is associated with the input noise and the inherent device noise of the active RIS
elements \cite{ref27}, 
and $\mathbf{n}\sim \mathcal{C}\mathcal{N}\left( 0,
\sigma^2\mathbf {I}_M \right)$ 
is the static noise
at the BS.

\section{  Channel Estimation}
In this section, we utilize the LMMSE method for obtaining the estimated aggregated  channel $\mathbf {\hat Q}$ based on the received pilot signals
in a single coherence interval 
 \cite{papazafeiropoulos2021intelligent}.
Each channel coherence interval 
spans $\tau_c$ time slots, of which $\tau$ time slots are utilized for channel estimation.
In each channel coherence interval, the 
 transmit pilot sequences from all users are mutually orthogonal and are simultaneously transmitted to the BS. Specifically,
the pilot sequence utilized by user $k$ is denoted as 
 ${\textbf{s}_{ k}} \in {{\mathbb C}^{{\tau} \times {1}}}$, then we have $\mathbf{S}={\left[\textbf{s}_{ 1},\textbf{s}_{ 2}...,\textbf{s}_{K}\right]}$, with $\mathbf{S}^H\mathbf{S}={\mathbf{I}}_K$. As a result, the pilot signals received at the BS can be given by
  \vspace{0.1cm}
 \begin{align}\label{接收到的导频信号}
 	{\mathbf{Y}_p}=\sqrt {\tau p} \mathbf{Q}\mathbf{S}^H + {\mathbf{H}}_2 \mathbf{A\Phi\Theta} \mathbf{V} +  \mathbf{N},
 \end{align}
where 
 $\mathbf{N} \in {\mathbb C}^ {  M\times \tau } $
 is the 
complex Gaussian noise
 matrix whose entries are i.i.d.
 with with zero mean and variance $\sigma ^{2}$
In addition, $\mathbf{V }
 \in {\mathbb C}^ {  N\times \tau } $
 is the thermal noise matrix where each entry is  modeled
 as a complex Gaussian random
 variable  with a mean of zero and a variance of $\sigma^2_e$.
 To eliminate the interference from other users, we multiply (\ref{接收到的导频信号}) by $\frac{\mathbf{s}_{ k}}{\sqrt {\tau p}}$ and leverage the orthogonality inherent in the pilot signals, obtaining the observation vector of user $k$ as below    
 \vspace{0.2cm}
 \begin{align}\label{观测信号}
 	{\mathbf{y}_p^k}=\frac{{\mathbf{s}_{ k}}}{\sqrt {\tau p}}{\mathbf{Y}_p} 
 	={\mathbf{q}}_k + \frac{\eta}{\sqrt {\tau p}}{\mathbf{H}}_2 \mathbf{\Phi\Theta} \mathbf{V}{\mathbf{s}_{k} }
 	+\frac{1}{\sqrt {\tau p}}\mathbf{N}{\mathbf{s}_{k}}.
 \end{align} 
\vspace{-0.1cm}

\begin{figure*}[hb]\label{估计的信道}
	\hrulefill
	\vspace{0.15cm}
		\setcounter{equation}{15}
	\begin{align}
		{{\hat {\mathbf q}}_k} ={{\bf{A}}_k}{\mathbf{y}_p^k} + {{\mathbf B}_k}
		=&
		{ 
			\underbrace
			{\eta \sqrt {{c_k}\delta {\varepsilon _k}}{{\bf{A}}_k} {{{\bf{\overline H}}}_2}{\bf{\Phi\Theta }}{{{\bf{\overline h}}}_{{k}}}}
			_{{\bf{\hat q}}_k^1}  
			+ 
			\underbrace 
			{\eta \sqrt {{c_k}\delta }{{\bf{A}}_k} {{{\bf{\overline H}}}_2}{{\bf{\Phi\Theta}} {\tilde {\mathbf h}_k}}^{\rm{}}}
			_{{\bf{\hat q}}_k^2}
			+
			\underbrace
			{\eta\sqrt {{c_k}{\varepsilon _k}}\mathbf A_k {{{{\tilde 
								{\mathbf H} }}}_2}{\bf{\Phi\Theta }}{{{\bf{\overline h}}}_{{k}}}}
			_{{\bf{\hat q}}_k^3}  
			+
			\underbrace 
			{\eta \sqrt {{c_k}}\mathbf A_k {{{{ {\tilde {\mathbf H} } }}}_2}{\bf{\Phi\Theta }}{{{\tilde{ \bf h}}}_k}}
			_{{\bf{\hat q}}_k^4} 
		} 
		\notag 
		\\
		+&
		\underbrace 
		{\eta\rho \sqrt {{c_k}\delta {\varepsilon _k}} {{{\bf{\overline H}}}_2}{\bf{\Phi }}{{{\bf{\overline h}}}_{{k}}}}
		_{{\bf{\hat q}}_k^5}  
		- 
		\underbrace 
		{  \eta\rho \sqrt {{c_k}\delta {\varepsilon _k}} \mathbf A_k               {{{\bf{\overline H}}}_2}{{\bf{\Phi }}{\overline {\mathbf h}_k}}^{\rm{}}}
		_{{\bf{\hat q}}_k^6}  			
		+{\frac{\eta}{\sqrt {\tau p}}\mathbf A_k{\bf{H}}_{{2}}}{{\bf{\Phi\Theta}} \mathbf{V}
			{\mathbf{s}} }_{ k}  
		+\frac{1}{\sqrt {\tau p}}\mathbf A_k\mathbf{N}{\mathbf{s}_{ k}}
		+\sqrt {{\gamma _k}} \mathbf A_k{{\tilde{\bf d} }_k}
			\label{estiamte q_k},
	\end{align}
	\vspace{-0.2cm}
\end{figure*} 	
	\begin{figure*}[hb]
	\hrulefill
	\vspace{0.15cm}
	\setcounter{equation}{16}
	\begin{align}
		{\rm NMSE}_k=&\frac{ {{\rm Tr}\{ {\rm{\mathbf  {MSE}}}_k\}}  }{{\rm Tr}\{{\rm{Cov}}\{ {\mathbf{q}}_k,\mathbf{q}_k\}\}}
		=\frac{  \left(\frac{\sigma^2}{\tau p}+\varpi+
			M\varpi\delta 
			\right)\left(\frac{\sigma^2}{\tau p}+\varpi\right)  (a_{k1}+a_{k2})
			+(\varpi\delta+\frac{\sigma^2}{\tau p}+\varpi)(a_{k2}^2+Ma_{k1}a_{k2})}
		{(a_{k2}+\frac{\sigma^2}{\tau p}+\varpi)
			\left\{(a_{k2}+\frac{\sigma^2}{\tau p}+\varpi)+M(a_{k1}+\varpi\delta)\right\}\left(a_{k1}+a_{k2}\right)	 }
			\label{NMSE_k},	
	\end{align}
	
\end{figure*}

Usually, the MMSE criterion can be employed to obtain optimal estimation of the channel
for user $k$.
While the MMSE estimator needs to know the moments of the channel as well as the fact that the channel follows or closely approximates a complex Gaussian distribution.
While the cascaded channel ${\mathbf g}_k$ is the product of two Gaussian variables and a random diagonal with Von Mises distribution in the considered system.
At the cost of little performance loss,
we adopt the LMMSE estimator due to its independence from the exact channel distributions.
Therefore, the LMMSE estimator is utilized to acquire the estimated channel $\mathbf {\hat Q}$.
By adopting MRC processing where  
 the received signal $\bf y$ is multiplied by the receive combining vector $\mathbf {\hat q}_k$, the expression for the $k$-th user of the received signal 
 can be formulated as
\begin{align}
	\setcounter{equation}{11}
	{{{r}}_{{k}}} = {\hat{ \mathbf q}}_k^H {\mathbf{y}}   =&
	 {\sqrt {p}  \,{\hat{ \mathbf q}}_k^H{{\bf{q}}_k}{x_k}} +  {\sqrt p \sum\limits_{i = 1,i \ne k}^K {{\hat{\mathbf q}}_k^H{{\mathbf{q}}_i}{x_i}} } \notag\\
	&+  {\eta\,{\hat{ \mathbf q}}_k^H{{\mathbf{H}}_{{2}}}{\mathbf{ \Phi\Theta} \pmb{\nu} }} 
	+ {{\hat{\mathbf q}}_k^H{\bf{n}}}, 
	 {k}\in\mathcal{K}\label{r_k},
\end{align}
where ${\hat{\mathbf q}_k}$ is the $k$-th column of $\hat{\mathbf{Q}}$.
Subsequently, we provide the necessary statistics for
 $\hat { \mathbf q}_k$ and 
 ${\mathbf{y}_p^k}$.
\vspace{0.2cm}

\begin{lemma}
The mean vectors and covariance matrices required for calculating the LMMSE estimator are expressed  as
	\begin{align}
		&\mathbb{E}\left\{ \mathbf{q}_k\right\}=\mathbb{E}\left\{ \mathbf{y}_p^k\right\}=
		{\eta \rho \sqrt {{c_k}\delta {\varepsilon _k}} {{{\bf{\overline H}}}_2}{\bf{\Phi }}{{{\bf{\overline h}}}_{{k}}}},\\
		&{\rm{Cov}}\{ \mathbf{q}_k,\mathbf{y}_p^k\} 
		= {\rm{Cov}}\{ \mathbf{q}_k,\mathbf{q}_k\}
		=a_{k1}\mathbf{a}_M\mathbf{a}_M^H+{a}_{k2}\mathbf{I}_M,\\
 &	{\rm{Cov}}\{ \mathbf{y}_p^k,\mathbf{y}_p^k\}   
 = m_{k} \mathbf{a}_{M}\mathbf{a}_{M}^H+ n_{k}{\mathbf I}_M, 
	\end{align}	
\end{lemma}
where $a_{k1}\!\!=\!{\Delta c_k\delta}\!\left\{(1-\rho^2)\varepsilon_k+1\right\}$,
$a_{k2}\!\!=\!\Delta c_k(\varepsilon _{k}\!+\!1)\!+\!\gamma _k$,
$m_k\!=
\!a_{k1}+\varpi\delta$ and 
$n_k\!=\!a_{k2}+\frac{\sigma^2}{\tau p}+\varpi$
with $\varpi\!=\!\frac{\Delta\beta\sigma_{e}^2}{\tau p(\delta+1)}$.

\vspace{0.1cm}
{\it \quad {Proof}} : See Appendix B.$\hfill\blacksquare$
\vspace{0.2cm}

\begin{theorem}
	The expression for the LMMSE estimate $\mathbf{\hat q}_k$ and
	the estimation's NMSE for $\mathbf{q}_k$ are represented by (\ref{estiamte q_k}) and
	(\ref{NMSE_k}) respectively
	at the bottom of this page, where
	\begin{align}
		\setcounter{equation}{17}
		&\mathbf{A}_k=\mathbf{A}_k^H=a_{k3}\mathbf{a}_M\mathbf{a}_M^H+a_{k4}\mathbf{I}_M,\\
		&\mathbf{B}_k=(\mathbf{I}_M-\mathbf{A}_k){\eta \sqrt {{c_k}\delta {\varepsilon _k}}}
		{{{\bf{\overline H}}}_2}{\bf{\Phi }}{{{\bf{\overline h}}}_{{k}}},\\
		&a_{k3}\!=\!\frac {   (\frac{\sigma^2}{\tau p}\!+\!\varpi)a_{k1} \!-\! \varpi\delta a_{k2}}
		{(a_{k2}\!+\!\frac{\sigma^2}{\tau p}\!+\!\varpi)^2\!+\! M(a_{k1}+\varpi\delta)(a_{k2}\!+\frac{\sigma^2}{\tau p}\!+\varpi)},\\
		&a_{k4}=\frac{a_{k2}}{a_{k2}+\frac{\sigma^2}{\tau p}+\varpi}.
	\end{align}	
	
	{\it \quad {Proof}} : See Appendix C.$\hfill\blacksquare$
	\vspace{0.1cm}
	
In detail, the first six terms of ${ \hat{\mathbf{q}}}_k$ are represented by 
${{\hat { \mathbf q}}_k^1} \sim {{\hat{ \mathbf q}}_k^6}$ respectively,
and we define 
${\underline{\hat  { \mathbf{q}}}_k}
=\sum_{w = 1}^6 { \hat {\mathbf {q}}_k^w} $
for the subsequent calculations.
	Meanwhile, we find that the phase noise does not manifest within (\ref{NMSE_k}), suggesting that the NMSE remains unaffected by the phase noise. Following this, we will explore some  insights related to  the NMSE.
	
		
%
\end{theorem}

\begin{corollary}
In the regime of both low and high pilot power-to-noise ratio,
 the asymptotic behaviors of NMSE are respectively formulated as
	\begin{align}
		\setcounter{equation}{21}
	\lim_{\frac{\sigma^2}{\tau p},\frac{\sigma_{e}^2}{\tau p} \to \infty}
\mathrm{NMSE}_k \to 1,\\
    \lim_{\frac{\sigma^2}{\tau p},\frac{\sigma_{e}^2}{\tau p} \to 0}
\mathrm{NMSE}_k \to 0.
\end{align}
\end{corollary}

Obviously, the value of NMSE 
which is between 0 and 1 
quantifies the  estimation error\cite{ref31}.
As we all know,
expanding the quantity of $N$ can yield a comparable outcome to enlarging the pilot sequence $\tau$, leading to a decrease in NMSE within the passive RIS-aided massive MIMO systems.
However, this approach proves ineffective for active RIS-aided massive MIMO systems since (\ref{NMSE_k}) is independent of $N$,
which suggests that the demand for reflecting elements in active RIS-aided systems is less critical compared to other systems aiming at enhancing channel estimation quality.
\begin{corollary}
When the RIS-BS channel is the
Rayleigh distributed, i.e., $\delta =0$,
we have
	\begin{align}
	&\mathrm{NMSE}_k=\frac{\frac{\sigma^2}{\tau p}+\Delta\frac{\sigma_e^2\beta}{\tau p} }
		{\Delta\beta\alpha _{k}+\gamma_k+
		\frac{\sigma^2}{\tau p}+\Delta\frac{\sigma_e^2\beta}{\tau p}}. 
	\end{align}
{\it\quad\quad {Proof}} : As $\delta =0$, we obtain $a_{k1}=0$, $a_{k2}=\Delta\beta\alpha_k +\gamma_k$ and $\varpi=\frac{\Delta \sigma_e^2\beta}{\tau p}$. 
The completion of the proof is achieved by substituting these results into (\ref{NMSE_k}). $\hfill\blacksquare$

Corollary 2 considers a certain scenario where considerable scatterers exist between the RIS and the BS. We also find the NMSE has a simple analytical expression where it is unaffected by both $M$ and $N$.
%
%
%
%
%
%
%
\end{corollary}

\section{Analysis Of The Achievable Rate}
In this section, we place emphasis on the derivation and analysis of the closed-form expressions for a lower bound of the 
achievable rate. 
Then, we investigate the power scaling laws in the system based on the 
theoretical
achievable rate. 
Additionally, we later substitute “achievable rate”
 for
“lower bound of the achievable rate” to streamline expressions
 in the content of this paper.


Before embarking on the achievable rate derivation, we first start with the calculation of the  amplification factor based on the amplification power  $P_A$.
\begin{lemma}
	The amplification power  $P_A$  can be calculated in a similar manner as \cite{liuxue2022_D2D}
	\begin{align}
		P_A =& \mathbb{E} \left\{ \left\| \sqrt {p}\mathbf{A\Phi\Theta H}_1
		\mathbf{x}+\mathbf{A\Phi\Theta 
			{\boldsymbol\nu} 
		} \right\| ^2 \right\}\notag\\ 
		=&\eta ^2N\big( \sum\nolimits_{k=1}^K{p\alpha _k+\sigma _{e}^{2}} \big).
	\end{align}

Although phase noise is present inside $P_A$, we observe that it does not have an impact on the  amplification power since its 
	characteristic function 
	has not been manifested.
	Then, given a specific amplification power, we can compute the factor as
	\begin{align}
		\eta = \sqrt{\frac{P_A} {N\big( \sum\nolimits_{k=1}^K{p\alpha _k+\sigma _{e}^{2}} \big)}}
	     \triangleq
		\sqrt {\frac{\Delta }{N}},
	\end{align}
where 	$\Delta 	= \sqrt{\frac{P_A} {\big( \sum\nolimits_{k=1}^K{p\alpha _k+\sigma _{e}^{2}} \big)}} $ for simplicity.
\end{lemma}

\subsection{Derivation of the Rate}
For tractable analysis, we take advantage of the use-and-then-forget (UatF) bound as in \cite{ref31} and \cite{2003hassibihow}
to 
characterize the lower bound of the ergodic
rate 
of the active RIS-aided massive MIMO system.
Specifically, by adding and subtracting the term  $\mathbb{E}\left\{
\hat{\mathbf q}_k^H \mathbf{q}_k
\right\}{x_k} $, (\ref*{r_k}) can be  reformulated  as
\vspace{0.15cm}
\begin{align}
	{{{r}}_{{k}}} = &\underbrace {{\sqrt p \,\mathbb{E}\left\{
		\hat{\mathbf q}_k^H \mathbf{q}_k
		\right\}}{x_k}}_{\textrm {Desired signal} }
		+\underbrace {\sqrt p \,\left(
	\hat{ \mathbf q}_k^H \mathbf{q}_k- {\sqrt p \,\mathbb{E}\left\{
			\hat{\mathbf q}_k^H \mathbf{q}_k
			\right\}}\right){x_k}}_{\textrm {Signal leakage} }
	\notag \\
	+& \underbrace {{\sqrt p \!\! \sum\limits_{i = 1,i \ne k}^K \! \!{{\hat{\mathbf  q}}_k^H{{\bf{q}}_i}{x_i}} }}_{ \textrm {Interference} }
	+ \underbrace {{\eta\,{\hat {\mathbf q}}_k^H{{\bf{H}}_{{2}}}{\mathbf{ \Phi\Theta} \pmb{\nu} }}
	+ 
		{{\hat{\mathbf q}}_k^H{\bf{n}}}
	}_{\textrm  {Noise}}.
\end{align}
\begin{figure*}[hb]\label{lower SINR}
	\hrulefill
	\vspace{0.1cm}
	\begin{align} 
		{\rm SINR}_k=\frac{p \, |   \mathbb{E} \left\{ \hat{\mathbf q}_k^H \mathbf{q}_k\right\}|^2    }
		{  p  \left(\mathbb{E}  \left\{| \hat {\mathbf q}_k^H \mathbf{q}_k|^2  \right\}
			-
			|   \mathbb{E} \left\{ \hat{\mathbf q}_k^H \mathbf{q}_k\right\}|^2 
			\right)+
			p \!\!{\sum\limits_{i = 1,i \ne k}^K}\!\! \mathbb{E}  \left\{ |\hat{\mathbf q}_k^H \mathbf{q}_i | ^2\right\} +
			\eta^2\sigma_e^2  \mathbb{E} \left\{  \left \| \hat{\mathbf q}_k^H\bf{H_2}  \bf{\Phi\Theta}\right \|^2     \right\}
			+ \sigma^2  \mathbb{E} \left\{  \left \| \hat {\mathbf q}_k\right \|^2     \right\}
		}
		\label{SINR_k}
		.
	\end{align}
\end{figure*}

Afterwards, we get the lower bound of the $k$-th user’s achievable rate as $\underline{{R}}_k={{\chi}\rm{log}_2(1+ {SINR}_k)}$, where $\chi=\frac{\tau_c-\tau}{\tau}$ denotes the factor that characterizes the utilization ratio of slots
per coherence block designated for transmission, and the effective ${\rm SINR}_k$ is presented in (\ref{SINR_k}) at the bottom of  this page.


\begin{theorem}
	The lower bound for the achievable rate of the $k$-th user is expressed as	$\underline{{R}}_k={\chi}{\rm log}_2(1+ {\rm SINR}_k)$, and the ${\rm SINR}_k$ is given by
	\begin{equation}
		\begin{aligned} 
			&{\rm SINR}_k \!\!=\!\! \frac{    pE^{ singal }_k(\bf \Phi)   }
			{{pE^{ leak }_k(\bf \Phi)}\!+\!  p{\sum\limits_{i = 1,i \ne k}^K}I_{ki}{(\bf{\Phi})} \!+\! E_k^{ noise}(\bf{\Phi}) 	}\label{sinr_k},
		\end{aligned}
	\end{equation}
\end{theorem}
where the specific expressions of  $E_k^{noise}(\bf{\Phi})$, ${E^{ signal }_k(\bf \Phi)}$,
$I_{ki}(\bf\Phi)$ and $E^{ leak }_k(\bf \Phi)$ are expressed as
 (\ref{E_phi_noise}), (\ref{E_phi_signal}), (\ref{I_phi_ki}) and (\ref{E_phi_leak})
respectively
in Appendix D.

\vspace{0.1cm}
{\it \quad Proof} : See Appendix D. $\hfill\blacksquare$
\vspace{0.1cm}

It is obvious that the effective ${\rm SINR}_k$  is only 
affected by slowly-varying statistical CSI.
Herein, perfect knowledge of the statistical CSI is assumed
at the BS as in \cite{ref16} and \cite{2022zhikangdapower}.
With the reduced computational complexity and feedback
overhead under the two-timescale framework, the phase shift design 
of the active RIS is  feasible with ease.
Besides, the lower bound of the  achievable sum rate is written as
	\begin{align} 
		&{{R}}={\chi}{\sum\limits_{k = 1}^K}{\rm{log}}_2(1+ {\rm SINR}_k).
	\end{align}

Next, we focus on analyzing the power scaling laws of the active RIS-aided system
given the presence of a huge number of the BS antennas $M$, which is important for
understanding the system's performance in large-scale configurations.
\subsection{Power Scaling Laws}

Within the massive MIMO system, transmit power is scaled down proportionally relative to the number of antennas in order to analyze the performance of the system
\cite{peng2022performance}.
Therefore,  the power scaling laws of the
active RIS-aided massive MIMO system
considering RIS-aided channels with different Rician factors
are discussed
 in this section.
%
For the sake of  clarity,
we use $a\geq0$ to represent the factor that determines the level of power scaling, while
$E_u$ denotes a constant value during the power scaling.

\begin{corollary}{
If the 
RIS-BS channel and the
user-RIS channel are all Rician distributed, i.e., $\delta > 0$ and $\varepsilon_k > 0,\forall k$, 
we find that the achievable rate    
approaches zero 						
as 
$M\rightarrow\infty$ and the transmit power $p$ of each user scales as $p={E_u/M^a}$.
}

{\it \quad Proof} : If $p={E_u/M^a}$ and $M\rightarrow\infty$, we retain the dominant part of $E^{ singal }_k(\bf \Phi)$, $E^{ leak }_k(\bf \Phi)$, $I_{ki}(\bf\Phi)$ and $E_k^{noise}(\bf{\Phi})$.
Then the achievable rate can be written as 
(\ref{R_k^{Ric,Ric}}) at the top of the next page, where
\begin{figure*}[ht]
\begin{align}
	\underline{{R}}_k^{(Ric,Ric)}\!\!
=\chi\mathrm{log_2}\left( 1+
\frac{ {E_uM^{2-a}} {\rm DS}^{r,1}_{k }    }
{  { {E_uM^{2-a}}   {\rm LE}^{r,1}_{k}   } 
	+
    {E_uM^{2-a}}\!{\sum\limits_{i = 1,i \ne k}^K}{\rm IN}^{r,1}_{i,k}
    +
    \sigma_e^2{M^2 {\rm TN}^{r,1}_{k}}
    +
    \sigma^2{M {\rm SN}^{r,1}_{k} }
}
		\right)\label{R_k^{Ric,Ric}},
\end{align}
	\hrulefill
\vspace{0.1cm}
\end{figure*}
\begin{align}
&{\rm DS}^{r,1}_{k}\!=\!
\frac{| f_k({\mathbf \Phi})|^4}{N^2}{\Delta^2}{c_k^2}{\delta^2}{\varepsilon_k^2}\rho^4, \\
&{\rm LE}^{r,1}_{k}\!=\!
\frac{|f_k({\mathbf \Phi})|^2}{N}{\Delta^2}{c_k^2}{\delta^2}{\varepsilon_k}\rho^2
\left\{ \varepsilon_k\left(1\!-\!\rho^2\right)+1
\right\}, 
%
%
\\
&{\rm IN}^{r,1}_{i,k}\!=\!\!
\frac{|f_k({\mathbf\Phi})|^2}{N}\!\!\Delta^2c_k \delta^2\!\varepsilon_k c_i\rho^2\!
\bigg\{
\!\varepsilon_i \!
\Big(
\!1\!-\!\rho^2\!\!+\! \! \frac{\rho^2|f_i({\mathbf\Phi})|^2}{N}
\!\Big)\!
\!\!+\!\!1\!\!
\bigg\},
\\
&{\rm TN}^{r,1}_{k}\!=\!
\frac{\beta}{ N  \left(\delta + 1\right)  }
\Delta^2|f_k({\mathbf\Phi})|^2
{c_k}{\delta^2}\varepsilon_k\rho^2, \\
&{\rm SN}^{r,1}_{k}\!=\!
\frac{|f_k({\mathbf\Phi})|^2}{N}\Delta c_k \delta \varepsilon_k\rho^2.
\end{align}

\vspace{0.1cm}
With an asymptotic behavior of $\mathcal O (M^{2-a})$, the desired signal term,
${E_uM^{2-a}} {\rm DS}^{r,1}_{k } $,
has a lower order than  the  thermal noise term, $\sigma_e^2{M^2 {\rm TN}^{r,1}_{k}}$, scaling as $\mathcal O (M^2)$ in (\ref{R_k^{Ric,Ric}}).
Consequently, the achievable rate converges to zero.$\hfill\blacksquare$

This scenario exemplifies that the power scaling laws of $M$ may not hold  when 
the amplified thermal noise is present.
Furthermore, we observe that the desired signal term is affected by the phase noise in the form of $\rho^2$, which is less than one according to Von Mises distribution.
However, if there is no phase noise, i.e.,  $v$ approaches infinity,  
the desired signal power can be enhanced as $\rho$ tends to one.
\end{corollary}

\begin{corollary}{
	If the user-RIS channel is Rayleigh distributed, i.e., $\varepsilon_k = 0$, $\forall k$, 
	and the 
		RIS-BS channel is Rician distributed, i.e., $\delta > 0$, 
	 the achievable rate of user $k$
		approaches zero,
		when 
		$p$ scales as $p={E_u/M^a}$ and $M\rightarrow\infty$.
		 } 
	
	{\it \quad Proof} : By retaining the dominant part of each term in the ${\rm SINR}_k$,
	 the achievable rate is given as (\ref{R_k^{Ric,Ray}}) at the top of this page, where the specific expressions of the dominant part are given by
\begin{figure*}[ht]
	\begin{align}
		\underline{{R}}_k^{(Ric,Ray)}\!\!
		=\chi\mathrm{log_2}\left( 1+
		\frac{ {E_uM^{2-a}}{\rm DS}^{r,2}_k   }
		{  { {E_uM^{2}}{\rm LE}^{r,2}_k  } 
			+
			{E_uM^{2}}\!{\sum\limits_{i = 1,i \ne k}^K}{\rm IN}_k^{r,2}
			+
			\sigma_e^2{M^{2+a} {\rm TN}^{r,2}_{k}}
			+
			\sigma^2{M {\rm SN}^{r,2}_{k} }
		}
		\right)\label{R_k^{Ric,Ray}},
	\end{align}
		\hrulefill
	\vspace{0.1cm}
\end{figure*}	
		\begin{align}
&{\rm DS}^{r,2}_{k}\!=\!
\left\{\Delta{c_k}\left( \delta{e_{k2}}+ e_{k1}\right) + \gamma_k e_{k1}
\right\}^2\label{DS^{r,2}_{k}},
 \\
&{\rm LE}^{r,2}_{k}\!=\!
\!\frac{\Delta^2\sigma_e^2\beta c_k}{\tau E_u\left(\delta+1 \right) }
\left\{
\delta^2e^2_{k2}\!+\! \frac{1}{N}
\left({2\delta e_{k1}e_{k2}}\!+\! e^2_{k1}\right)\!
\right\}\label{LE^{r,2}_{k}},
\\
&{\rm IN}^{r,2}_{i,k}\!=\!
\frac{\Delta^2\sigma_e^2\beta c_i}{\tau E_u\left(\delta+1 \right) }
\left\{
\delta^2e^2_{k2}\!+\! \frac{1}{N}
\left({2\delta e_{k1}e_{k2}}\!+\! e^2_{k1}\right)\!
\right\}\label{IN^{r,2}_{i,k}},\\
&{\rm TN}^{r,2}_{k}=
\frac{\Delta^2\delta\sigma_e^2\beta^2 e_{k2} }{\tau E_u\left(\delta+1 \right)^2 }
\big(
\delta e_{k2} +\frac{2e_{k1}}{N}
\big)\label{TN^{r,2}_{k}},
\\
&{\rm SN}^{r,2}_{k}\!=\!
\Delta c_k\delta e_{k2} + \Delta c_k e_{k1} + \gamma_k e_{k1}\label{SN^{r,2}_{k}}
,
	\end{align}
where $e_{k1}$ and $e_{k2}$ scale as $\mathcal O (M^{-a})$.
Hence, 
as a function of $M$, 
(\ref{DS^{r,2}_{k}})-(\ref{TN^{r,2}_{k}})
 behave asymptotically as $\mathcal O (M^{-2a})$, and 
 (\ref{SN^{r,2}_{k}}) 
 scales as  $\mathcal O (M^{-a})$. 
As a result,
the order of 
${E_uM^{2-a}}{\rm DS}^{r,2}_k$
is $\mathcal O (M^{2-3a})$ which is lower than that of
the thermal noise term, 
$\sigma_e^2{M^{2+a} {\rm TN}^{r,2}_{k}}$,
with an order of $\mathcal O (M^{2-a})$. Therefore, the achievable rate also tends to zero.$\hfill\blacksquare$
\end{corollary}

Likewise, the power scaling law “$1/M^a$" is not applicable when the RIS-BS channel is Rician distributed. 

\begin{corollary}
	{
	If the RIS-BS channel and the user-RIS channel  
	are all  Rayleigh distribution. With $p={E_u/M^a}$,  the achievable rate of user $k$
	approaches zero as $M\rightarrow\infty$.}

	{\it \quad Proof} : When $\delta=0$,
	we have ${a_{k1} }=0$, ${a_{k2}}=\Delta\beta\alpha_k+\gamma_k$, ${a_{k3} }=0$ and  ${e_{k3}}={e^2_{k1} }$.
	By substituting these values into the expressions in Theorem 2, 
and ignoring the terms with lower order as $M\rightarrow\infty$,
the achievable rate is written as  (\ref{R_k^{Ray,R}}) at the top of the next page,
\begin{figure*}[ht]
	\begin{align}
		\underline{{R}}_k^{(Ric,R)}\!\!
		=\chi\mathrm{log_2}\left( 1+
		\frac{ {E_uM^{2-a}}{\rm DS}^{r,3}_k   }
		{  { {E_uM^{2}}{\rm LE}^{r,3}_k  } 
			+
			{E_uM^{2}}\!{\sum\limits_{i = 1,i \ne k}^K}{\rm IN}_k^{r,3}
			+
			\sigma_e^2{ M \left(M{\rm TN}^{r,3}_{k,1}+M^a{\rm TN}^{r,3}_{k,2}\right) }
			+
			\sigma^2{M {\rm SN}^{r,3}_{k} }
		}
		\right)\label{R_k^{Ray,R}},
	\end{align}
	\hrulefill
	\vspace{0.1cm}
\end{figure*}	
where the dominant terms are given as
\begin{align}
&{\rm DS}^{r,3}_{k}\!=\!
e^2_{k1}
\left( \Delta\beta\alpha_{k}  +  \gamma_k\right) ^2\label{DS^{r,3}_{k}},
\\
&{\rm LE}^{r,3}_{k}\!=\!
 \frac{1}{N}{e^2_{k1}}\Delta^2\frac{\sigma^2_e}{\tau E_u}\beta^2\alpha_k, \\ 
&{\rm IN}^{r,3}_{i,k}\!=\!
\frac{1}{N}\Delta^2{e^2_{k1}}
 \beta^2\alpha_{i} \frac{\sigma^2_e}{\tau E_u},      \\
&{\rm TN}^{r,3}_{k}=
	\underbrace 
{ \frac{1}{N}{e^2_{k1}}\Delta^2\beta^2\alpha_k}_{{\rm TN}^{r,3}_{k,1}}
+
	\underbrace 
{\Delta{e^2_{k1}}\beta
\frac{\Delta \sigma^2_e\beta + \sigma^2 }{\tau E_u}}_{{\rm TN}^{r,3}_{k,2}}
\label{TN^{r,3}_{k}}
,
\\
&{\rm SN}^{r,3}_{k}\!=\!
{e_{k1}}\left( \Delta\beta\alpha_k+\gamma_k   \right)\label{SN^{r,3}_{k}},	
\end{align}				
where $e_{k1}$ scales as $\mathcal O (M^{-a})$.
Therefore, 
(\ref{DS^{r,3}_{k}})-(\ref{TN^{r,3}_{k}})
behave asymptotically as $\mathcal O (M^{-2a})$, and  
(\ref{SN^{r,3}_{k}})
scales as  $\mathcal O (M^{-a})$.

However, the order of the thermal noise term depends on the factor $a$. 
If $a\in(0,1)$, it scales with $\mathcal O (M^{2-2a})$. Otherwise, its asymptotic behavior  is $\mathcal O (M^{1-a})$.
Since the order of the desired signal term,
${E_uM^{2-a}}{\rm DS}^{r,3}_k$, 
scaling as 
$\mathcal O (M^{2-3a})$
is always lower than that of the the thermal noise term, 
$\sigma_e^2{ M \left(M{\rm TN}^{r,3}_{k,1}+M^a{\rm TN}^{r,3}_{k,2}\right) }$,
the achievable rate reaches zero.$\hfill\blacksquare$

\end{corollary}

Corollary 5 reveals that the power scaling law fails to hold for the considered system 
in the case where  the  channel between the RIS and the BS  is Rayleigh distributed.

From Corollary 3 to Corollary 5, we draw some conclusions.
Firstly, it is observed that the thermal noise introduced by the inherent structural characteristics \cite{ref6} of the active RIS unavoidably appears in the dominant denominator for any given $a$. 
With an increase in $M$, the desired signal power gradually diminishes, whereas the power of the thermal noise amplified  remains unaffected. 
Secondly, 
the phase noise exerts an impact on the ${\rm SINR}_k$, distinctly reflected in the desired signal power containing $\rho^2$ in the scenario where all the channels are Rician distributed.
While if $\delta=0$ or $\varepsilon_k=0$, the impact disappears.
Thirdly, as the achievable rate in (\ref{R_k^{Ric,Ray}}) and (\ref{R_k^{Ray,R}}) is independent of $\bf\Phi$, any RIS phase shift yields the equivalent achievable rate when $\delta=0$ or $\varepsilon_k=0$. Hence, designing the phase shifts is unnecessary 
if there are many scatterers between the BS and the RIS or between the users and the RIS.

Nevertheless, in large system parameter configuration,
the power scaling laws become less critical.
We can still leverage the active RIS to promote the system performance. This is because in practical deployment, the quantity of $M$ is finite and what we need is simply to reduce the transmit power while still maintaining the communication. 

Similar to the case of $M$, we can know the power scaling laws related to $N$, where 
the achievable rate is zero as $N$ approaches infinity.
The simulation results of 
power scaling laws with $N$
will partly show in Section \uppercase\expandafter{\romannumeral 6}.  
In addition, increasing $N$ can  improve the  active RIS-aided system's performance to a certain extend. However, considering the structure of active RISs, blindly increasing $N$  will affect the allocation of power as more power being consumed to initiate RIS.
One the other hand, high amplified power and transmit power can obviously help improve system performance. Therefore, it is necessary to consider the trade-off between the quantity of $N$ and the power allocation, which needs further research work.

\section{Design Of The RIS Phase Shifts}
In this section, the optimization of phase shifts based on statistical CSI is considered.
Specifically, as fairness among multiple users need to be ensured,
this optimization aims to maximize minimum user achievable rate. 
Herein, we formulate the optimization problem as
\vspace{0.1cm}
\begin{align}
	&\mathop{\max}_{\mathbf{\Phi }} \quad \mathop{\min}_{k\in\mathcal{K}}   \quad
	\underline{{R}}_k(\bf \Phi),  
	\\
	&\,\,\text{s.t.} \quad  0\leq  \theta_n \textless\,  2\pi, \forall n,
\end{align}
where 
$\underline{{R}}_k(\bf \Phi)$  is given 
in Theorem 2, 
$\bf \Phi$ is the phase shift matrix, and $\theta_n$ is the phase shift at the $n$-th element of the active RIS.
Considering the complexity of $\underline{{R}}_k(\bf \Phi)$, resolving the problem with conventional methods becomes challenging. 
By viewing the RIS phase shifts as the genes of a population,
we apply the GA to tackle this optimization problem.
Specifically, the fundamental concept of GA applied to RIS-aided systems involves treating the  phase shift of each RIS element  
as the genetic code of a chromosome. 
Through iterative updates to these genetic codes, the population evolves. 
This eventually results in the RIS phase shifts being set to the most optimal codes identified in the last generation.
Following that, we further discuss the specific procedural steps.

   {\it 1)  Population initialization:} The initial population comprises $N_p$ individuals. 
   The chromosome of each individual is composed of $N$ genes,
   where the $n$-th gene $\theta_n$ is generated within $[0, 2\pi)$.

   {\it 2)  Fitness evaluation and scaling:} 
   In the current population, the fitness evaluation function for individual $i$ 
   with chromosome ${\bf \Phi}^i$ is defined
   as  $f_i=\mathop{\min}\limits_{k}\underline{{R}}_k({\bf \Phi}^i)$.
Subsequently, the raw fitness scores derived from the fitness function undergo a scaling process to be adjusted within a range suitable for the selection function.
   Specifically, we sort the $N_p$ individuals in descending order based on their raw fitness scores
   representing an individual's adaptation to the environment.
   An individual's position 
  within the sorted scores 
   corresponds to its rank, 
   offering insight into its relative fitness level among the population.
   Individual $i$ with rank $r_i$ has scaled score proportional to 
   $\frac{1}{\sqrt{r_i}}$,
  and the sum of the scaled values across the population equals the population size of parents for the next generation, which means that 
 the expected fitness can be calculated as
\vspace{0.3cm}
   \begin{align}
   f_i^{e}=\frac{1/\sqrt{r_i}}{\begin{matrix} \sum_{i=1}^{N_p} 
   		1/\sqrt{r_i}
   	\end{matrix}}.
\end{align}
 \vspace{0.1cm}
 
  {\it 3)  Selection:} Then, we preserve $N_e$ individuals with high expected fitness as elites for the next generation. Meanwhile, the stochastic universal sampling method 
  given in Algorithm 1 is utilized twice to select 2$N_c$ parents for crossover.
Specifically, we create a line where each parent is associated with a segment of the line, each segment's length is proportional to its expected fitness.  
The algorithm progresses along the line in uniform steps,
assigning a parent from the segment it lands on at each step.
   \begin{algorithm}[H]
   	\caption{Stochastic universal sampling}
   	\begin{algorithmic}
   		\STATE 1:Input the step size $p_{step}=1/N_c$, and set $sum$ = $0$;
   		\STATE 2:Randomly generate the initial pointer $p_0\in[0,p_{step}]$;
   		\STATE 3: {\textbf {for}} j = 1 : $N_c$ {\bf do}
   		\STATE 4: \hspace{0.5cm}{\bf{ for}} i = 1 : $N_p$ {\bf do}
   		\STATE 5:  \hspace{0.8cm}$sum$ = $sum$ + $f_i^{e}$;
   		\STATE 6:  \hspace{0.8cm} {\textbf {if}} \, $p_0<$ $sum$ {\textbf {then}} 
   		\STATE 7:  \hspace{1.2cm}Individual $i$ becomes a parent and then break;
   		\STATE 8:  \hspace{0.92cm}{\textbf {end if}}
   		\STATE 9:  \hspace{0.65cm}{\textbf {end for}}
   		\STATE 10: \hspace{0.35cm} Set $sum$ = $0$, $p_0=p_0+ip_{step}$;
   		\STATE 11:\hspace{0.0cm}{\textbf {end for}}
   	\end{algorithmic}
   	\label{alg1}
   \end{algorithm}
   
 {\it 4) Crossover:} After selecting parents, we combine two individuals to produce a crossover offspring for the subsequent generation. 
 The procedure for crossover is outlined in Algorithm 2.
   \begin{algorithm}[H]
	\caption{Single-point Crossover}
	\begin{algorithmic}
		\STATE 1:Input the selected parents
		\STATE 2: {\textbf {for}} i = 1 : $N_c$ {\bf do}
	    \STATE 3: \hspace{0.5cm}Select the ($2i-1$)-th and $2i$-th as the parents; 
	    \STATE 4:  \hspace{0.5cm}Generate a random  integer $n$,  $n\in[1,N]$;
		\STATE 5:  \hspace{0.5cm}Choose vector entries with numbers $\leq n$ from the 	\STATE     \hspace{0.95cm}first parent. Select vector entries with numbers $\textgreater n$
		 \STATE     \hspace{0.95cm}from the other parent. The child has a chromosome
		 \STATE     \hspace{0.95cm}${\rm diag}$
		 $
		 \left\{
		 e^{j\theta_1^{2i-1}}, ..., e^{j\theta_n^{2i-1}}, e^{j\theta_{n+1}^{2i}}, ..., e^{j\theta_N^{2i}}      
		 \right\}
		    $;
		\STATE 6:\hspace{0.14cm}{\textbf {end for}}
	\end{algorithmic}
	\label{alg2}
\end{algorithm}
 {\it 5) Mutation:} Afterwards, $N_m$ mutated offspring are generated as in Algorithm 3, where  the $i$-th offspring is reproduced  by mutating each gene 
 of $i$-th parent
 under  
 a probability
 $p_m$.
\begin{algorithm}[H]
	\caption{Mutation}
	\begin{algorithmic}
		\STATE 1: {\textbf {for}} i = 1 : $N_m$ {\bf do}
		\STATE 2: \hspace{0.5cm}{\textbf {for}} n = 1 : $N$ {\bf do}
		\STATE 3:  \hspace{0.8cm}Generate $c$ randomly from (0,1);
		\STATE 4:  \hspace{0.85cm}{\textbf {if}} \, $c<$ $p_m$ {\textbf {then}} 	
		\STATE 5:    \hspace{1cm}The $n$-th entry $e^{j\theta_{n}^{i}}$ in chromosome of   
		 parent $i$
		\STATE \hspace{1.4cm}mutates to a value randomly selected from a 
    	\STATE \hspace{1.4cm}uniform distribution in $[0, 2\pi)$;
		\STATE 6:  \hspace{0.86cm}{\textbf {end if}}
		\STATE 7:  \hspace{0.51cm}{\textbf {end for}}
		\STATE 8:\hspace{0.14cm}{\textbf {end for}}
	\end{algorithmic}
	\label{alg3}
\end{algorithm}
After performing these procedures, we obtain the next generation composed of $N_e$ elites, $N_c$ offspring generated through crossover and $N_m$ offspring produced by mutation. The algorithm will stop when it reaches the maximum number of iterations $100N$ or the average change
of the raw fitness is lower than $10^{-4}$.
As a result, the GA algorithm outputs the chromosome of the individual with the highest fitness in the current population. 
   
\section{Numerical Results}
In this section, we present numerical simulations that  are conducted to evaluate the influence of important parameters on the performance of an active RIS-aided massive MIMO system. 
With a setup similar to 
 \cite{zhi2022two-timescale}, we posit an assumption where $K=8$ users are evenly distributed along a semicircle with the active RIS as the center  and 
  a radius $r_{U\!R}$ of 20 m.
The distance from the RIS to the BS is $d_{RB}$ = $700$ m. While the distance between  user $k$ and the BS is calculated as $d_k^{U\!B}$=
$\sqrt{\left(d_{RB}-r_{U\!R}{\rm cos}(\frac{\pi}{9}k)\right)^2
	+\left(r_{U\!R}{\rm sin}(\frac{\pi}{9}k)\right)^2 }$. 
Owing to the long distance between users and the BS, coupled with the presence of obstacles,
the direct links suffer severe attenuation
compared to the RIS-aided links, i.e.,
the path-loss exponent of the direct links exceeds that of the RIS-aided links.
Therefore, the large-scale fading
coefficients for user-RIS channel, RIS-BS channel and user-BS channel  are modeled respectively as $\alpha_{k}\!=\!10^{-3}r_{U\!R}^{-2}$, $\beta\!=\!10^{-3}d_{RB}^{-2.8}$ and $\gamma_k \!=\!10^{-3}{d_k^{U\!B}}^{-4.2},\forall k$.
Each coherence block, as our assumption, comprises $\tau_c \! = \!196$  symbols, out of which 
$\tau \! = \!8$ symbols are utilized for channel estimation.
Furthermore, we set 
the static noise power as 
$\sigma^2\!=\!-104 $ dBm and the thermal noise power as $\sigma_{e}^2 \!= \!- 70 $ dBm. 
Moreover, 
we consider systems
with passive RIS and those without
RIS in the absence of phase noise for comparison. 
Meanwhile, for the phase noise at the active RIS, we have the concentration parameter $v$ = 2.

 Differing from passive RIS
 that  has zero direct-current (DC) power consumption, 
 active RIS reflecting elements need the suitable
 DC biasing power to operate\cite{ref29}. As a result, 
 the total power consumption of the uplink active RIS-aided system
 is given by
 \begin{align}
 	 P_{total}= \sum\limits_{k = 1}^Kp + 
 	 P_{cir}
 	 +\xi^{-1}P_A,
 \end{align}
 where 
 $P_{cir}=N\left(P_{SC}+P_{DC}\right)$ is the circuit power. 
 Specifically, the power consumption of the switch and control (SC) circuit at each reflecting element is  
 denoted by 
 $P_{SC}$ and the power consumption of DC is expressed as $P_{DC}$.
Besides, 
 $\xi$ = 0.8 is the amplifier efficiency. 
 It is noted that if $ P_{total}\leq N\left(P_{SC}+P_{DC}\right)$, the active RIS does not work, and a similar situation applies to passive RIS when $P_{p} \leq NP_{SC}$ where $P_{p}$ is the total power consumption in the passive RIS-aided system.
Herein, we  take into account that the $P_{SC}\!= \!-10$ dBm and $P_{DC}\!=\! -5$ dBm. Unless explicitly stated otherwise, we assume $P_{total} \!=\! 20$ dBm for all systems. 
Furthermore,  an equal power
splitting scheme is adopted, where $\xi^{-1}P_A= \sum\limits_{k = 1}^Kp$.
Additionally, the lines labeled as “Simulation" follow from (\ref{SINR_k}),
whereas those marked as “Theory" are acquired based on (\ref*{sinr_k}).

 
To begin with, we evaluate the NMSE and  neglect the power consumption of RIS circuits
for 
the
simplicity of analysis.
\begin{figure}[!t]
	\centering
	\includegraphics[width=3.5in]{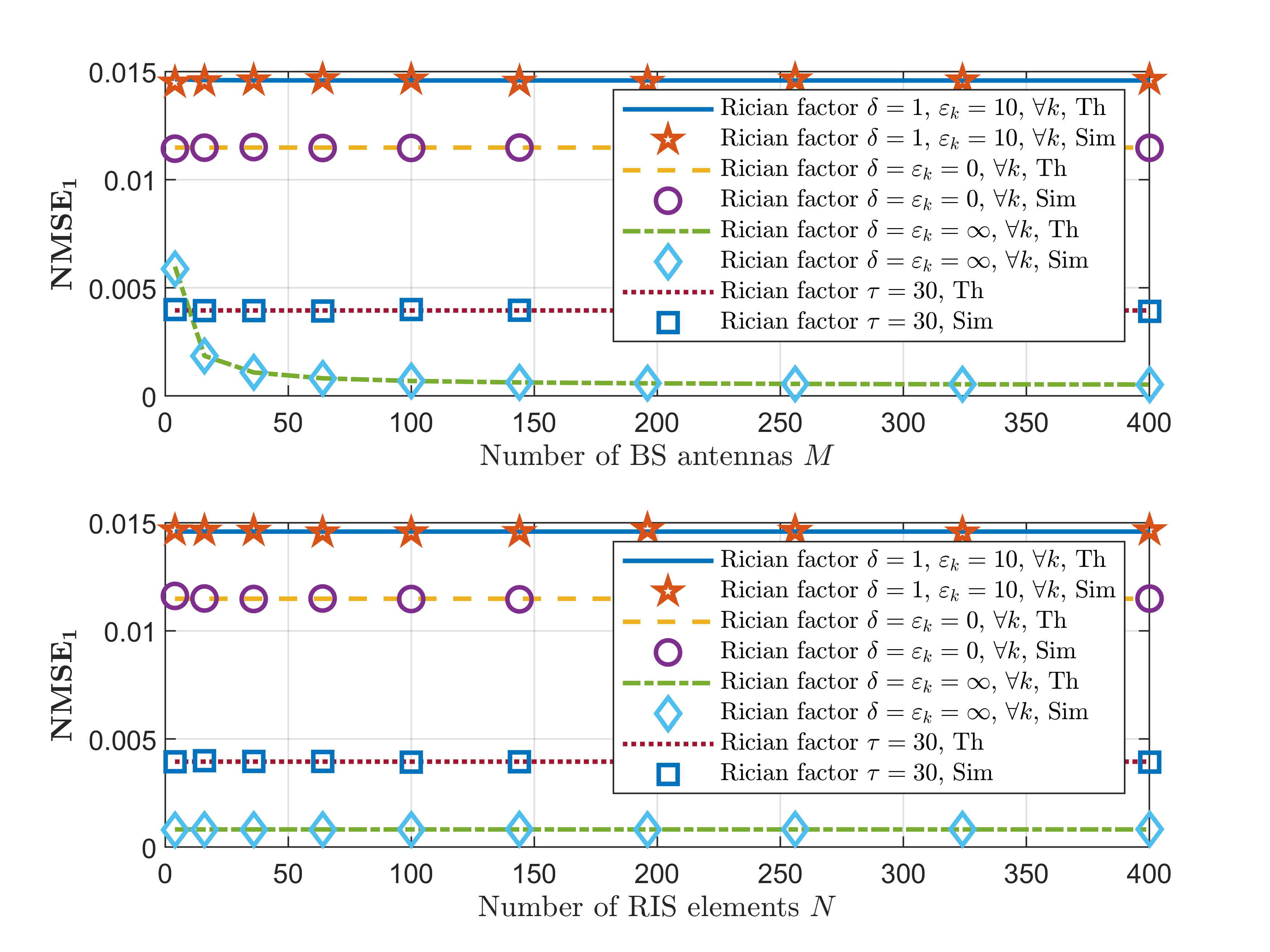}
	\caption{ NMSE of user 1 versus the number of antennas $M$ and the number of RIS elements $N$.}
	\label{fig_2}
\end{figure}
In Fig. 2, we illustrate the NMSE of user 1 against the number of 
$M$ and $N$.
Generally speaking, the NMSE is not particularly sensitive to changes in $M$ and $N$, except that it is a decreasing function as $M$ increases when $\delta = \varepsilon_k \!\rightarrow\!\infty$.
Also, through the extension of the pilot signals to 30, a decrease in NMSE is noted.
Furthermore, we observe that, unlike passive RIS, the NMSE of active RIS-aided system 
 is not dependent on
 $N$. This implies that there is no need to sacrifice deployment costs for smaller NMSE. This further corroborates the findings in Corollary 1.

In Fig. 3, we present the relationship between the total power and achievable
sum rate.
Initially, the achievable sum rate of the passive RIS-aided system slightly exceeds that of the active counterpart 
primarily. This is because of the active RIS's additional power requirements for startup, reducing the available power for amplification and transmission.
However, as the total power increases, the achievable rate of the active RIS-aided system surpasses that of the passive counterpart. 
Furthermore, the  GA algorithm 
we discuss 
in Section \uppercase\expandafter{\romannumeral 5} contributes to achieving  better performance when applied to the phase shifts of  both active and  passive RIS, compared to random fixed phase shifts.
\begin{figure}[!t]
	\centering
	\includegraphics[width=3.5in]{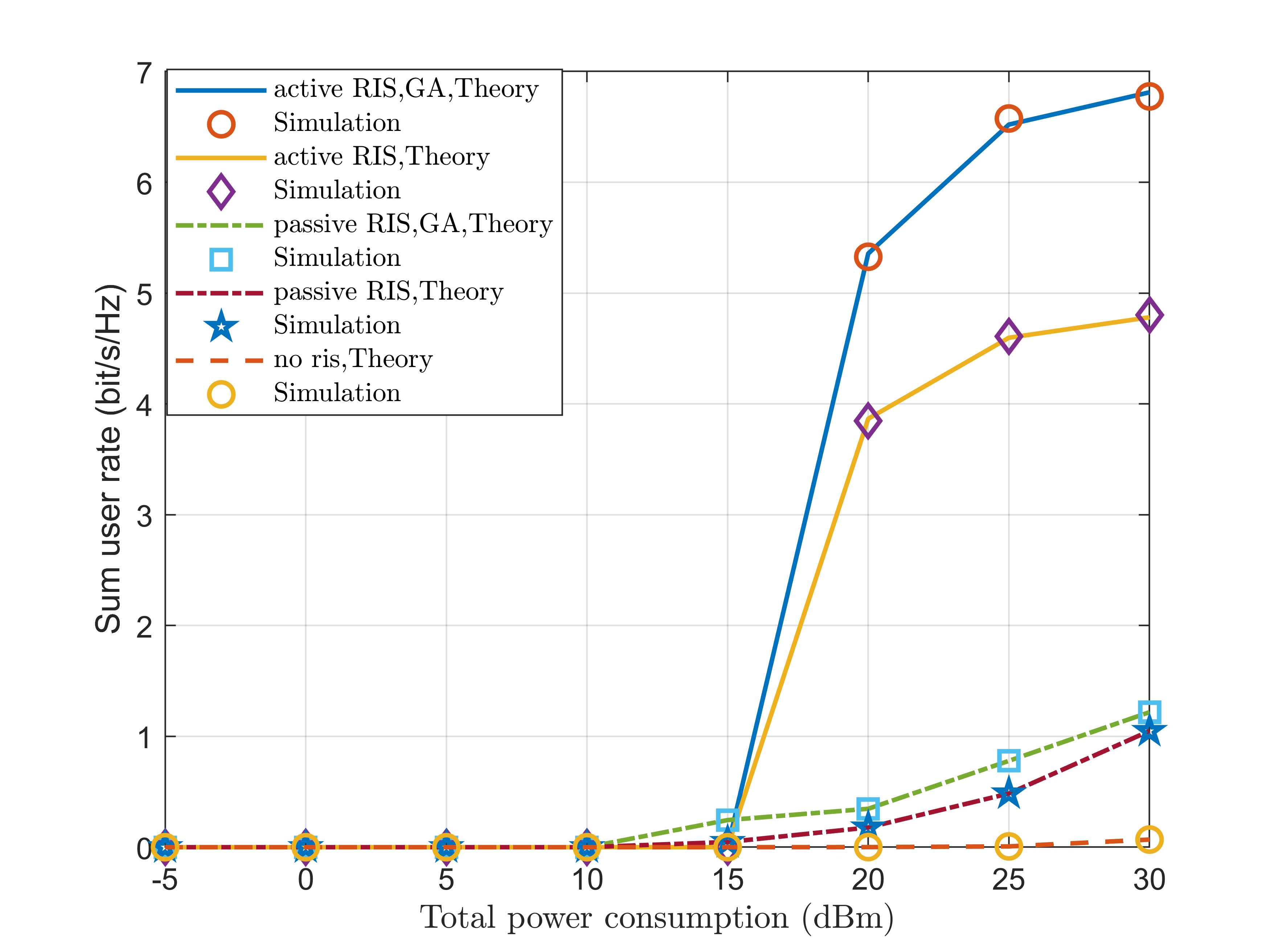}
	\caption{ Uplink achievable sum rate versus total power consumption.}
	\label{fig_3}
\end{figure}

Fig. 4 illustrates how the achievable sum rate exhibits variations corresponding to the number of BS antennas $M$ when utilizing the optimized phase shifts. Remarkably, as $M$ grows, the sum rate of active RIS-aided system exhibits an increase and its trend behaves similarly to that of passive RIS-aided system. 
Despite the rate approaching saturation as $M$ increases, the active RIS-aided system consistently maintains a significantly higher rate compared to the passive 
counterpart.
Then we use the example of Rician channels to illustrate intuitively why the power scaling laws are not applicable to active RIS-aided system. Specifically,
the transmit power is  scaled down by $p\! =\!\frac{E_u}{M}$, 
$p \! =\!\frac{E_u}{\sqrt M}$ and $p \!=\!\frac{E_u}{M^2}$ with $E_u \!= \!10$ dBm respectively.  From Fig. 4, we observe that as $M$ increases, the rate shows a progressive decrease instead of remaining constant, due to the rapid power scaling down. 
If we further increase $M$, the rate is expected to eventually converge to zero. 

\begin{figure}[!t]
	\centering
	\includegraphics[width=3.5in]{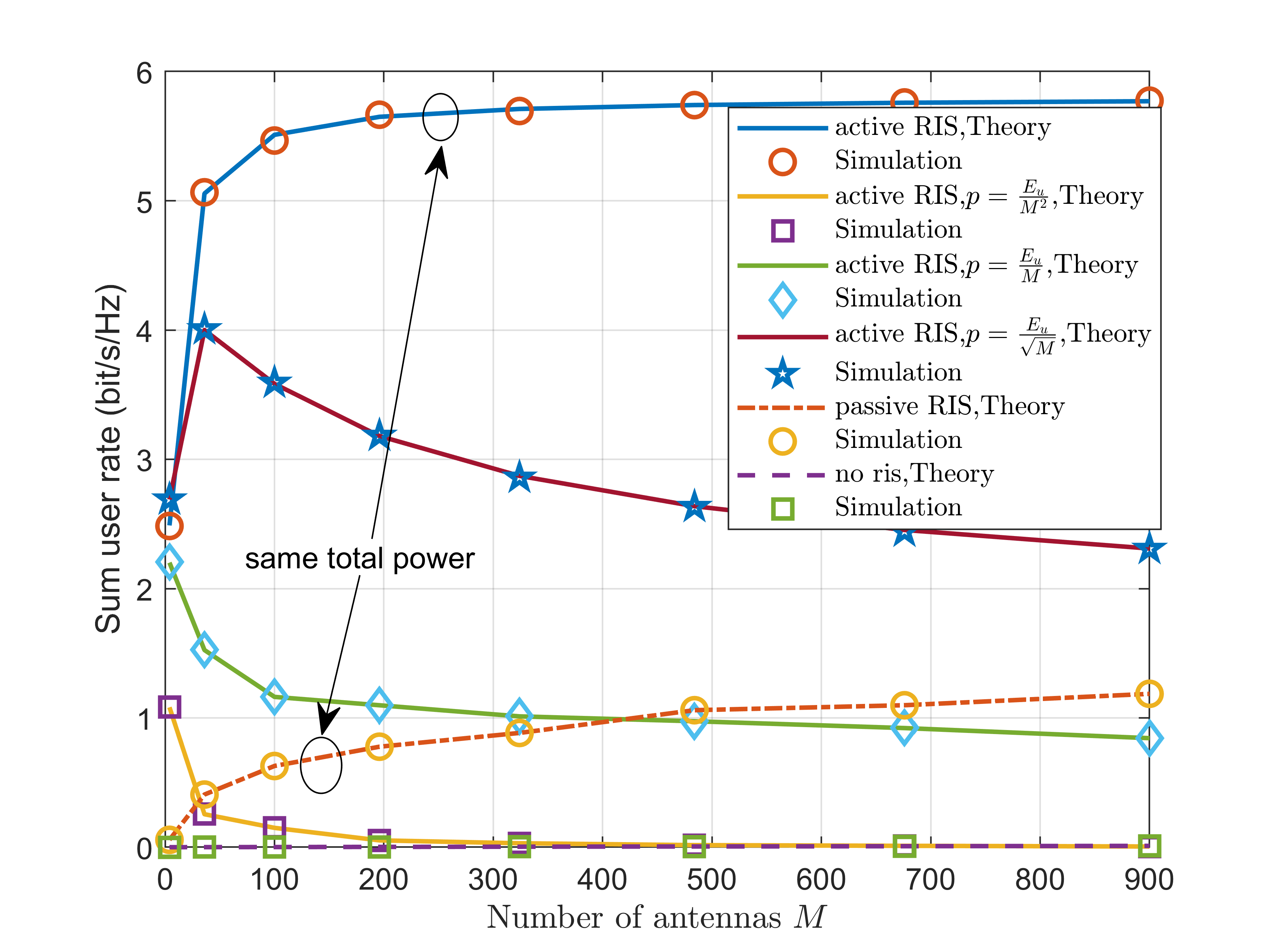}
	\caption{ Uplink achievable sum rate versus the number of antennas $M$.}
	\label{fig_4}
\end{figure}

Fig. 5 shows the achievable sum rate with respect to the number of elements $N$ with optimized phase shifts.
Similarly, the achievable sum rate of the active RIS-aided system  markedly surpasses than that of the passive counterpart.
If we neglect the power consumption of active RIS circuits,
the rate experiences an increase as $N$ grows, but it quickly reaches saturation. 
If we scale down the transmit power proportionally to $\frac{1}{N}$ and $\frac{1}{\sqrt N}$, we still find that the power scaling laws are not applicable to active RIS-aided system,
in fact, worsens the situation.
In particular, the reduction in the sum rate occurs initially because the transmit power is scaled down, while the thermal noise remains unaffected. This results in the desired power being relatively small compared to the thermal noise.
When $N$ increases, the RIS requires huge power support as the circuit power consumption has a linear relationship with $N$,
which results in the majority of power being allocated to the circuits in active RIS elements. 
Therefore, in large scale configuration of $N$, the achievable rate immediately approaches zero 
due to insufficient power for the circuits in the RIS elements.
Hence, in practice, the number of $N$ should be smaller
to better leverage the role of active RIS within the system.

\begin{figure}[!t]
	\centering
	\includegraphics[width=3.5in]{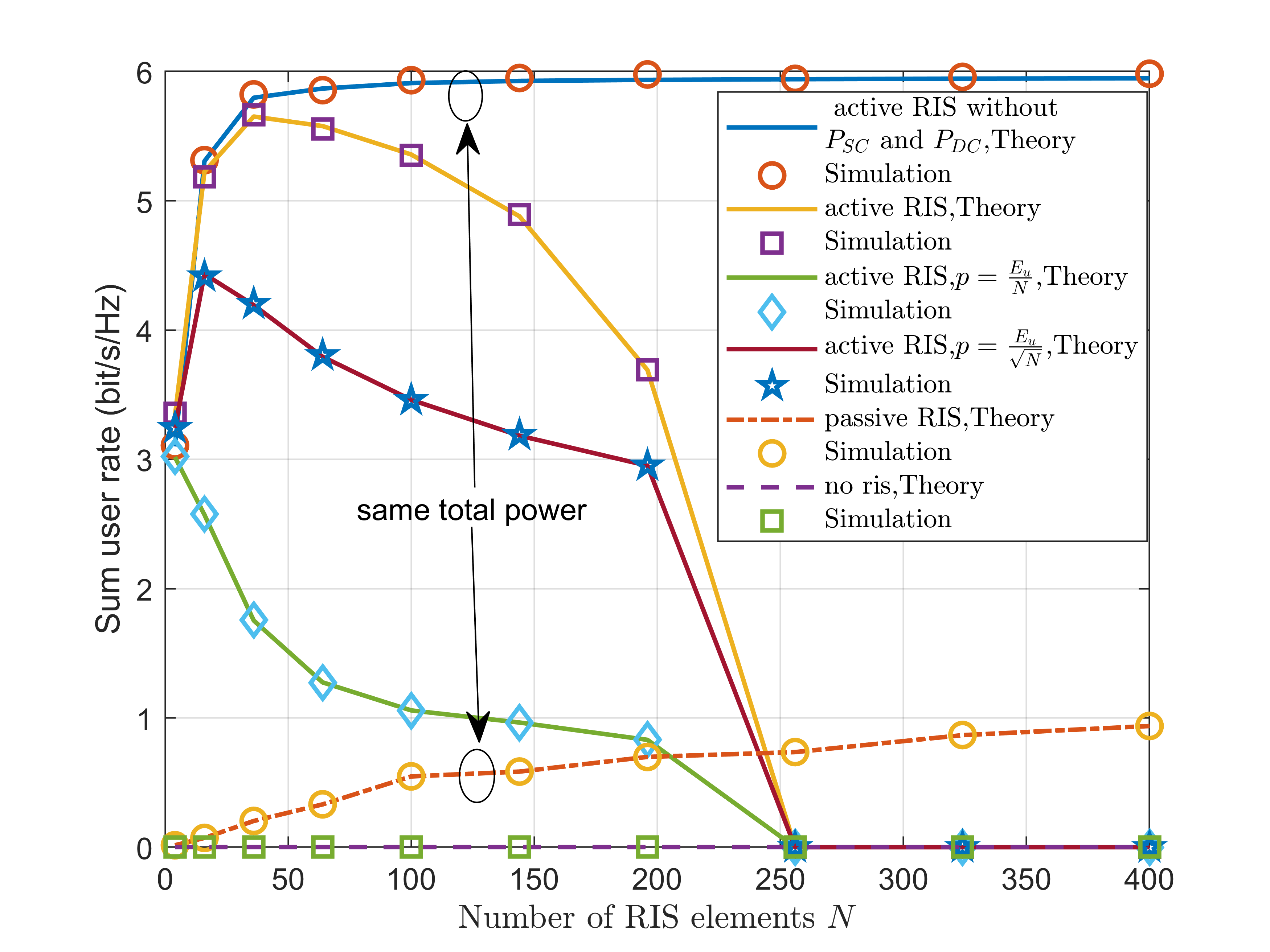}
	\caption{ Uplink achievable sum rate versus the number of RIS elements $N$.}
	\label{fig_5}
\end{figure}
\begin{figure}[!t]
	\centering
	\includegraphics[width=3.5in]{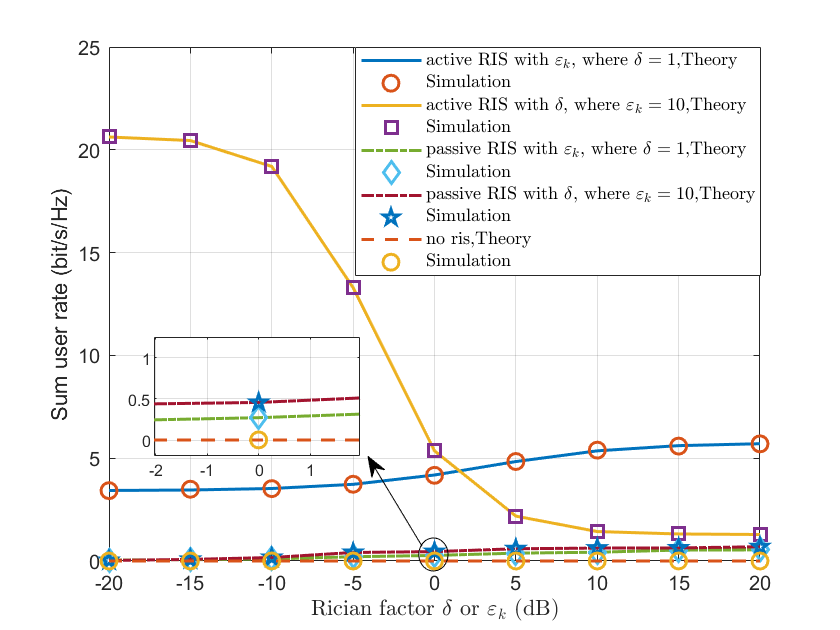}
	\caption{ Uplink achievable sum rate versus rician factor $\delta$ or $\varepsilon_k$.}
	\label{fig_6}
\end{figure}
In Fig. 6,
the achievable sum rate experiences a decline as $\delta$ increases, while it demonstrates an upward trend with the rise of $\varepsilon_k$ in the active RIS-aided system.
%
This is because when $\delta$ is small, the RIS-BS channel is rich scattering, which increases the spatial multiplexing gains, thereby improving the performance of the system. 
While with increasing $\delta$,  the rank of the RIS-BS channel converges to one, which is not conducive to spatial multiplexing for multiple users, leading to  a low achievable rate \cite{wu2021intelligent}.
On the other hand, as $\varepsilon_k$ increases, the user-RIS channels become LoS-dominant, providing better beamforming gain \cite{zheng2021double}.
Also, it's interesting to find that in the case of relatively low total power, increasing $\delta$ can actually improve the achievable sum rate in the passive RIS-aided system. 

\section{Conclusion}
In this work, we delved into the closed-form expressions of an active RIS-aided massive MIMO system, taking into account the influence of the phase noise, within the framework of a two-timescale design scheme.
Specifically, we considered Rician fading for RIS-aided channels and Rayleigh fading for the direct links.
Then we utilized the LMMSE channel estimator for the 
aggregated channels with the MRC detector to receive signals.
The UatF bound of achievable rate 
was derived in the closed-form, depending only on statistical CSI. 
Subsequently, we investigated the power scaling laws, revealing that scaling the transmit power by $1/M^a$ or $1/N^a$ does not allow for maintaining a non-zero achievable rate in the active RIS-aided system.
Furthermore, we performed a GA-based method to optimize the phase shifts of the active RIS. 
Our results validated the consistently superior performance of an active RIS-aided system
even when subjected to phase noise,
compared to the passive counterpart under different system parameters. 

{\appendices
\section*{{ \large A}PPENDIX { \large A}\\
{ \large P}ROOF OF {\large S}OME {\large U}SEFUL {\large R}ESULTS
}	

First, we review some useful lemmas used in the following computations.
\begin{lemma}
Considering 
the deterministic  matrix
  ${\bf{\overline h}}_{k}  {\overline {\mathbf h}}_k^H \in {{\mathbb C}^{{N} \times {N}}}$ and the random phase noise matrix 
$\mathbf{\Theta}\in {{\mathbb C}^{{N} \times {N}}}$, we have
\begin{align}
&\mathbb{E}
\left\{
{\mathbf \Theta}{\bf{\overline h}}_{k}  {\overline {\mathbf h}_k^H} {{\mathbf \Theta}}^H	\right\} 
=
\rho^2{\bf{\overline h}}_{k}  {\overline {\mathbf h}_k^H} +(1-\rho^2){\rm \mathbf I}_N.
\end{align}	

{\it \quad Proof} :   We define $[ {\rm {\bf X}}  ]_{mn}$ as the entry in 
the $m$-th row and $n$-th column of matrix ${\rm {\bf X}}$. 
Since the PDF of $\tilde{\theta_n }$ is symmetric, 
there's
$\mathbb{E}\left\{e^{j\tilde{\theta }_m}\right\}\!=\!
\mathbb{E}\left\{e^{-j\tilde{\theta }_n}\right\}\!=\!\rho $. 
When $m \neq n$, we can get
\begin{align}
\left[\mathbb{E}
\left\{	{\mathbf \Theta}{{\mathbf{\overline h}}_{k}}  {\overline {\mathbf h}_k^H }{{\mathbf \Theta}}^H	\right\}\right]_{mn}\!&=\!
\left[
{\bf{\overline h}}_{k}  {\overline {\mathbf h}_k^H}
\right]_{mn}\!\!
\mathbb{E}\left\{e^{j\tilde{\theta }_m-{j\tilde{\theta }_n}}\right\} \notag \\
&=\rho^2
\left[
{\bf{\overline h}}_{k}  {\overline {\mathbf h}_k^H}
\right]_{mn}.
\end{align}	
With 
${\rm diag}\left\{{\bf{\overline h}}_{k}{\overline{\mathbf h}_{k}^H}\right\}={\mathbf I}_N$,  we obtain that
	\vspace{0.25cm}
\begin{align}
\mathbb{E}
\left\{
{\mathbf \Theta}{\bf{\overline h}}_{k}  {\overline {\mathbf h}_k^H} {{\mathbf \Theta}}^H	\right\}
&= 
\rho^2{\bf{\overline h}}_{k}  {\overline {\mathbf h}}_k^H +(1-\rho^2){\rm diag}\left\{{\bf{\overline h}}_{k}{\overline{\mathbf h}_{k}^H}\right\} \notag
\\
&=
\rho^2{\bf{\overline h}}_{k}  {\overline {\mathbf h}}_k^H +(1-\rho^2){\rm \bf I}_N.
\end{align}

Then, we complete the proof.$\hfill\blacksquare$
\vspace{5pt}
\end{lemma}		

\begin{lemma}
As in \cite{zhi2022two-timescale},
	we take into account a matrix ${\rm {\bf X}}\in  {{\mathbb C}^{{M} \times {N}}}$, whose entries are i.i.d. with zero mean and $v_x$ variance. With a deterministic matrix ${\rm {\bf D}}\in  {{\mathbb C}^{{N} \times {N}}}$, we have
	\begin{align}
		\mathbb{E}\left\{ {\rm {\bf X}}{\rm {\bf D}}{\rm {\bf X}}^H  \right\} 
		=v_x {\rm Tr}\left\{ {\rm {\bf D}}\right\}\mathbf {I}_M.
	\end{align}	



For both deterministic matrices  ${\rm {\bf C}}\in  {{\mathbb C}^{{M} \times {M}}}$ and ${\rm {\bf D}}\in  {{\mathbb C}^{{N} \times {N}}}$, 
if ${\rm {\bf C}}={\rm {\bf C}}^H $, we have 
  	\begin{align}
  	&\mathbb{E}\left\{ {{{\tilde{\mathbf H}}}_2^H}{\rm {\bf C}} {{{\tilde{\mathbf H}}}_2}
  	{\rm {\bf D}}{{{\tilde{\mathbf H}}}_2^H}{\rm {\bf C}} {{{\tilde{\mathbf H}}}_2}
  	 \right\} \notag \\
  	&= {\rm Tr}\left\{ {\rm {\bf D}}\right\} {\rm Tr}\left\{ {\rm {\bf C}^2}\right\}\mathbf {I}_N +
  |  \rm {Tr}\left\{ {\rm {\bf C}}\right\}|^2\rm {\bf D}.    
  \end{align}	
	
\end{lemma}	

	\section*{{\large A}PPENDIX {\large B}\\
	{ \large P}ROOF OF {\large L}EMMA {\large 1}
	}
	\vspace{-0.5cm}
Referring to $\mathbf y^k_p$ where  ${{{\tilde{ \mathbf H}}}_2}$, ${{{\tilde{\bf h}}}_k}$, ${{{\tilde{\mathbf d}}}_k}$, $\mathbf{V}$ and $\mathbf N$ are mutually independent with zero-mean and 
$\mathbb{E}\left\{e^{j\tilde{\theta }_m}\right\}=\rho$,  
we have
\begin{align}
	\mathbb{E}\left\{ \mathbf{y}_p^k\right\}&=
	\mathbb{E}\left\{ \mathbf{q}_k\right\}\!+\! 
	\frac{\eta}{\sqrt {\tau p}}\mathbb{E}\left\{\mathbf{H}_2 \mathbf{\Phi\Theta} \mathbf{V}{\mathbf{s}_{k} }\right\}	\notag
	\!+\!\frac{1}{\sqrt {\tau p}}\mathbb{E}\left\{ \mathbf{N}{s}_{k} \right\}\notag	\\ 
&= {\eta\rho \sqrt {{c_k}\delta {\varepsilon _k}} {{{\mathbf{\overline H}}}_2}{\mathbf{\Phi    }}{{{ \mathbf{\overline h}}}_{{k}}}}.
\end{align}	
	
As a result, the covariance matrix between
 the channel 
 $\mathbf q_k$ and 
 the observation vector 
 ${\mathbf{y}_p^k}$ can be expressed respectively as
\begin{align}
	&{\rm{Cov}}\{ \mathbf{q}_k,\mathbf{y}^k_p\} =  \,
	\mathbb{E}\left\{ ( \mathbf{q}_k-\mathbb{E}\left\{ \mathbf{q}_k \right\}   )
	(\mathbf{y}_p^k-\mathbb{E}\left\{ \mathbf{y}_p^k \right\} )^H
	\right\}\notag
	\\
	&=\mathbb{E}\left\{
 \left( \mathbf{q}_k-\mathbb{E}\left\{ \mathbf{q}_k \right\}\right) \notag
 \right.\\& \left. 	\times
 \left(
 \mathbf{q}_k+ \frac{\eta}{\sqrt {\tau p}}\mathbf{H}_2\mathbf{\Phi\Theta} \mathbf{V}\mathbf{s}_{ k}+\frac{1}{\sqrt  {\tau p}} \mathbf{N}\mathbf{s}_{ k}-\mathbb{E}\left\{ \mathbf{q}_k \right\}
\right)^H
	\right\} \notag \\
	&=\mathbb{E}\left\{ ( \mathbf{q}_k-\mathbb{E}\left\{ \mathbf{q}_k \right\}   )
	(\mathbf{q}_k-\mathbb{E}\left\{ \mathbf{q}_k \right\} )^H \notag 
	\right\}\\
	&={\rm{Cov}}\{ \mathbf{q}_k,\mathbf{q}_k\}. 
\end{align}	
\vspace{0.2cm}
and we have
\begin{align}
		{\rm{Cov}}\{ \mathbf{y}_p^k,\mathbf{q}_k\}& =({\rm{Cov}}\{\mathbf{q}_k,\mathbf{y}_p^k\} )^H \notag \\ 
		&=({\rm{Cov}}\{ \mathbf{q}_k,\mathbf{q}_k\})^H
		={\rm{Cov}}\{ \mathbf{q}_k,\mathbf{q}_k\}.
\end{align}	

\vspace{0.36cm}
With the definition of $\mathbf q_k$, ${\rm{Cov}}\{ \mathbf{q}_k,\mathbf{q}_k\}$ can be written as 
\begin{align}
&	{\rm{Cov}}\{ \mathbf{q}_k,\mathbf{q}_k\} \notag \\
&	=\mathbb{E} \Big \{ \big(
	{\eta \sqrt {{c_k}\delta {\varepsilon _k}} {{{\bf{\overline H}}}_2}{\bf{\Phi\Theta }}{{{ \bf{\overline h}}}_{{k}}}}		
	+{\eta \sqrt {{c_k}\delta } {{{\bf{\overline H}}}_2}{\bf{\Phi\Theta }}
		{{{\tilde{ \bf h}}}_k} }	\notag \\
	&\qquad \ \  +{\eta \sqrt {{c_k}{\varepsilon _k}   } {{{\tilde{ \bf H}}}_2}{\bf{\Phi\Theta }}
		{{{\overline{ \bf h}}}_k} }	
	+{\eta \sqrt {{c_k}  } {{{\tilde{ \bf H}}}_2}{\bf{\Phi\Theta }}
		{{{\tilde{ \bf h}}}_k} }	+ \sqrt {{\gamma _k}} {{\tilde{\bf d} }_k}\notag \\
	&\qquad \ \  -{\eta\rho \sqrt {{c_k}\delta {\varepsilon _k}} {{{\mathbf{\overline H}}}_2}{\mathbf{\Phi    }}{{{ \mathbf{\overline h}}}_{{k}}}}
	\big)\!\!\times\!\!
\left( \mathbf{q}_k\!-\!{\eta\rho \sqrt {{c_k}\delta {\varepsilon _k}} {{{\bf{\overline H}}}_2}{\bf{\Phi}}{{{ \bf{\overline h}}}_{{k}}}} \!\!  \right) ^H\!\!
	\Big \}.
\end{align}	

\vspace{0.3cm}

Then, the first term in ${\rm{Cov}}\{ \mathbf{q}_k,\mathbf{q}_k\}$ can be calculated as 
\begin{align}
&\mathbb{E}\left\{ 
\left(
{\eta \sqrt {{c_k}\delta {\varepsilon _k}} {{{\bf{\overline H}}}_2}{\bf{\Phi\Theta }}{{{ \bf{\overline h}}}_{{k}}}}
\right)
\left( \mathbf{q}_k-{\eta\rho \sqrt {{c_k}\delta {\varepsilon _k}} {{{\bf{\overline H}}}_2}{\bf{\Phi}}{{{ \bf{\overline h}}}_{{k}}}}   \right) ^H
\right\}  \notag\\
&=\mathbb{E}\left\{ 
\left(
{\eta \sqrt {{c_k}\delta {\varepsilon _k}} {{{\bf{\overline H}}}_2}{\bf{\Phi\Theta }}{{{ \bf{\overline h}}}_{{k}}}} 
\right) \notag \right.\\& \left.
\quad \times\left( 
\eta \sqrt {{c_k}\delta {\varepsilon _k}}
 {\overline {\mathbf h}}_k^H {{\mathbf \Theta}}^H
	{\bf{\Phi}}^H   {\overline {\mathbf H}}_2^H 
-{\eta\rho \sqrt {{c_k}\delta {\varepsilon _k}} {{{\bf{\overline H}}}_2}{\bf{\Phi}}{{{ \bf{\overline h}}}_{{k}}}}   \right) ^H
\right\}  \notag\\
&={\eta^2  {{c_k}\delta {\varepsilon _k}} {{{\bf{\overline H}}}_2}{\bf{\Phi }}
	\mathbb{E}
	 \left\{
	{\mathbf \Theta}{\bf{\overline h}}_{k}  {\overline {\mathbf h}}_k^H {{\mathbf \Theta}}^H	\right\}
	{\bf{\Phi}}^H   {\overline {\mathbf H}}_2^H} \notag\\ 
&-{\eta^2\rho  {{c_k}\delta {\varepsilon _k}} {{{\bf{\overline H}}}_2}{\bf{\Phi }}
	\mathbb{E}
	\left\{
{{\mathbf \Theta}}	\right\}{\bf{\overline h}}_{k}{\bf{\overline h}}_{k}^H
	{\bf{\Phi}}^H   {\overline {\mathbf H}}_2^H} \notag \\ 
&=\eta^2(1-\rho^2){{c_k}\delta {\varepsilon _k}}N\mathbf{a}_{M}\mathbf{a}_{M}^H.
\end{align}

The remaining terms are obtained by using a similar procedure. Therefore we have

\begin{align}
&\mathbb{E}\left\{ 
\left(
{\eta \sqrt {{c_k}\delta} {{{\bf{\overline H}}}_2}{\bf{\Phi\Theta }}{{{ \tilde{\bf h}}}_{{k}}}}
\right)
\left( \mathbf{q}_k-{\eta\rho \sqrt {{c_k}\delta {\varepsilon _k}} {{{\bf{\overline H}}}_2}{\bf{\Phi}}{{{ \bf{\overline h}}}_{{k}}}}   \right) ^H
\right\}  \notag\\
&=\mathbb{E}\left\{ 
\left(
{\eta \sqrt {{c_k}\delta} {{{\bf{\overline H}}}_2}{\bf{\Phi\Theta }}{{{ \tilde{\bf h}}}_{{k}}}}
\right)
\left( 
{\eta \sqrt {{c_k}\delta} {{{\bf{\overline H}}}_2}{\bf{\Phi\Theta }}{{{ \tilde{\bf h}}}_{{k}}}}
 \right) ^H
\right\}  \notag\\
&=\eta^2c_k\delta N \mathbf{a}_{M}\mathbf{a}_{M}^H,
\\
	&\mathbb{E}\left\{ 
	\left(
	{\eta \sqrt {{c_k}{\varepsilon _k}} 
	{{{\tilde{\bf H}}}_2}{\bf{\Phi\Theta }}{{{ \overline{\bf h}}}_{{k}}}}
	\right)
	\left( \mathbf{q}_k-{\eta\rho \sqrt {{c_k}\delta {\varepsilon _k}} {{{\bf{\overline H}}}_2}{\bf{\Phi}}{{{ \bf{\overline h}}}_{{k}}}}   \right) ^H
	\right\}  \notag\\
	&=\mathbb{E}\left\{ 
	\left(
	{\eta \sqrt {{c_k}{\varepsilon _k}} 
		{{{\tilde{\bf H}}}_2}{\bf{\Phi\Theta }}{{{ \overline{\bf h}}}_{{k}}}}
	\right)
	\left(
		{\eta \sqrt {{c_k}{\varepsilon _k}} 
		{{{\tilde{\bf H}}}_2}{\bf{\Phi\Theta }}{{{ \overline{\bf h}}}_{{k}}}}
	  \right) ^H
	\right\}  \notag\\
	&=\eta^2c_k\varepsilon_kN {\mathbf I}_M,
	\\
&	\mathbb{E}\left\{
	{\eta \sqrt {{c_k}  } {{{\tilde{ \bf H}}}_2}{\bf{\Phi\Theta }}
		{{{\tilde{ \bf h}}}_k} }
	\left( \mathbf{q}_k-{\eta\rho \sqrt {{c_k}\delta {\varepsilon _k}} {{{\bf{\overline H}}}_2}{\bf{\Phi}}{{{ \bf{\overline h}}}_{{k}}}}   \right) ^H		
			\right\}\notag\\
&=\mathbb{E}\left\{
	{\eta \sqrt {{c_k}  } {{{\tilde{ \bf H}}}_2}{\bf{\Phi\Theta }}
	{{{\tilde{ \bf h}}}_k} }
		\left(
	{\eta \sqrt {{c_k}  } {{{\tilde{ \bf H}}}_2}{\bf{\Phi\Theta }}
	{{{\tilde{ \bf h}}}_k} }
\right) ^H
\right\}\notag\\
&=\eta^2c_kN {\mathbf I}_M,
\\
	&\mathbb{E}\left\{ 
	\sqrt {{\gamma _k}} {{\tilde{\bf d} }_k}
	\left(  \mathbf{q}_k-
	{\eta\rho \sqrt {{c_k}\delta {\varepsilon _k}} {{{\bf{\overline H}}}_2}{\bf{\Phi}}{{{ \bf{\overline h}}}_{{k}}}}
	\right) ^H
	 \right\}\notag\\
	&=\mathbb{E}\left\{ 
	\sqrt {{\gamma _k}} {{\tilde{\bf d} }_k}
	\sqrt {{\gamma _k}} {{\tilde{\bf d} }^H_k}
	\right\}
	=\gamma _k{\mathbf I}_M,
	\\
	&\mathbb{E}\left\{ 
	\left(
		-{\eta\rho \sqrt {{c_k}\delta {\varepsilon _k}} {{{\mathbf{\overline H}}}_2}{\mathbf{\Phi    }}{{{ \mathbf{\overline h}}}_{{k}}}}
	\right)
	\left(  \mathbf{q}_k-
	{\eta\rho \sqrt {{c_k}\delta {\varepsilon _k}} {{{\bf{\overline H}}}_2}{\bf{\Phi}}{{{ \bf{\overline h}}}_{{k}}}}
	\right)^H
	\right\}\notag\\
	&=\mathbb{E}\bigg\{ 
	\left(
	-{\eta\rho \sqrt {{c_k}\delta {\varepsilon _k}} {{{\mathbf{\overline H}}}_2}{\mathbf{\Phi    }}{{{ \mathbf{\overline h}}}_{{k}}}}
	\right)\left(
	{\eta\sqrt {{c_k}\delta {\varepsilon _k}} {{{\mathbf{\overline H}}}_2}{\mathbf{\Phi\Theta    }}{{{ \mathbf{\overline h}}}_{{k}}}}
	\right)^H\notag \\
&\qquad	+
	\left(
	-{\eta\rho \sqrt {{c_k}\delta {\varepsilon _k}} {{{\mathbf{\overline H}}}_2}{\mathbf{\Phi    }}{{{ \mathbf{\overline h}}}_{{k}}}}
	\right)\left(
	-{\eta \sqrt {{c_k}\delta {\varepsilon _k}} {{{\mathbf{\overline H}}}_2}{\mathbf{\Phi  \Theta  }}{{{ \mathbf{\overline h}}}_{{k}}}}
	\right)^H \!
	\bigg\}\notag \\
	&=0.
\end{align}	

Thus, the covariance matrix can be obtained as
\begin{align}
&{\rm{Cov}}\{ \mathbf{q}_k,\mathbf{q}_k\}
\notag \\
&=\eta^2(1-\rho^2){{c_k}\delta {\varepsilon _k}}N\mathbf{a}_{M}\mathbf{a}_{M}^H
+\eta^2c_k\delta N \mathbf{a}_{M}\mathbf{a}_{M}^H
\notag \\
&\quad +\eta^2c_k\varepsilon_kN {\mathbf I}_M+\eta^2c_kN {\mathbf I}_M+\gamma _k{\mathbf I}_M+0\notag \\
&\triangleq a_{k1} \mathbf{a}_{M}\mathbf{a}_{M}^H+ a_{k2}{\mathbf I}_M,
\end{align}	
where we define $a_{k1}=
\Delta c_k\delta \left\{\varepsilon_k\left(1-\rho^2\right)+1\right\}$, and 
$a_{k2}=\Delta c_k\left(\varepsilon_k+1\right) + \gamma_k$.

\vspace{0.25cm}
Therefore, ${\rm{Cov}}\{ \mathbf{y}^k_p,\mathbf{y}^k_p\}$ can be expressed as
\vspace{0.2cm}
 \begin{align}
 	&{\rm{Cov}}\{ \mathbf{y}^k_p,\mathbf{y}^k_p\} =  \,
 	\mathbb{E}\left\{ ( \mathbf{y}^k_p-\mathbb{E}\left\{ \mathbf{y}^k_p \right\}   )
 	(\mathbf{y}_p^k-\mathbb{E}\left\{ \mathbf{y}_p^k \right\} )^H
 	\right\}\notag
 	\\
 	&=\mathbb{E}\left\{
 	\left( \mathbf{q}_k \!-\!\mathbb{E}\left\{ \mathbf{q}_k \right\}\right)
 	\left( \mathbf{q}_k \!-\!\mathbb{E}\left\{ \mathbf{q}_k \right\}\right)^H \!
 	\right\} 
 	\!+\!
 	 \frac{1}{\tau p}\mathbb{E}\!\left\{{\mathbf N\mathbf{s}_{ k} }
 	\mathbf{s}_{ k} ^H{\mathbf N}^H\!
 		\right\}
 	 \notag \\
 	&+\mathbb{E}\left\{  
 \left(\frac{\eta}{\sqrt {\tau p}}\mathbf{H}_2\mathbf{\Phi\Theta} 
 \mathbf{V}\mathbf{s}_{ k} \right)	
  \left(\frac{\eta}{\sqrt {\tau p}}\mathbf{H}_2\mathbf{\Phi\Theta} 
  \mathbf{V}\mathbf{s}_{ k} \right)^H
 	\right\} \notag  \\
 	&={\rm{Cov}}\{ \mathbf{q}_k,\mathbf{q}_k\}\!+\! \frac{\sigma^2}{\tau p}{\mathbf I}_M
 	\!+\!\frac{\eta^2\sigma_e^2\beta}{\tau p\left(\delta+1\right)}
 	\mathbb{E}\left\{ \!\delta
 	{   \mathbf  {\overline H}   }_2{   \mathbf  {\overline H}   }_2^H
 	\!+\!{{{\tilde{ \mathbf H}}}_2} {{{\tilde{ \mathbf H}}}^H_2} \!
 	 \right\}
 	\notag 
 	 \\
 	&={\rm{Cov}}\{ \mathbf{q}_k,\mathbf{q}_k\}\!+\!
 	   \frac{\sigma^2}{\tau p}{\mathbf I}_M \!+\!
 	   \frac{\eta^2\sigma_e^2\beta N\delta}{\tau p\left(\delta+1\right)}\mathbf{a}_{M}\mathbf{a}_{M}^H
 	 \!+\! \frac{\eta^2\sigma_e^2\beta N}{\tau p\left(\delta+1\right)}\!{\mathbf I}_M
 		\notag \\
&=\left( a_{k1}+\varpi\delta
\right)\mathbf{a}_{M}\mathbf{a}_{M}^H 
+
\left( a_{k2}+\frac{\sigma^2}{\tau p}+\varpi
\right)\mathbf{I}_{M}^H \notag \\
&\triangleq m_{k} \mathbf{a}_{M}\mathbf{a}_{M}^H+ n_{k}{\mathbf I}_M
\label{Cov{y_p^k,y_p^k}},
 \end{align}	
where $m_{k}= a_{k1}+\varpi\delta$ and 
$ n_{k}= a_{k2}+\frac{\sigma^2}{\tau p}+\varpi$.

\vspace{1cm}
	\section*{{\large A}PPENDIX {\large C}\\
	{ \large P}ROOF OF {\large T}HEOREM {\large 1}
	}
Based on the
observation vector  $\mathbf{y}_p^k$,
the LMMSE estimate of the channel $\mathbf{q}_k$ is formulated as in   \cite{ref35}, and we have
 \begin{align}
	{{\hat{ \mathbf q}}_k} &=\mathbb{E}\left\{ \mathbf{q}_k \right\} \notag \\
 &+{\rm{Cov}}\{\mathbf{q}_k,\mathbf{y}^k_p\}
 	{\rm{Cov}}^{-1}\,\{ \mathbf{y}^k_p,\mathbf{y}^k_p\} 
 		(\mathbf{y}_p^k-\mathbb{E}\left\{ \mathbf{y}_p^k \right\} ).
 \end{align}	
For the
${\rm{Cov}}^{-1}\{ \mathbf{y}^k_p,\mathbf{y}^k_p\}$ as in \cite{ref35},
we can get
 \begin{align}
	{\rm{Cov}}^{-1}\{ \mathbf{y}^k_p,\mathbf{y}^k_p\} 
	= n^{-1}_{k}{\mathbf I}_M-
	\frac{m_{k}\left(n_{k}\right)^{-2} }{1+Mm_{k}n_{k}^{-1}}\mathbf{a}_{M}\mathbf{a}_{M}^H,
\end{align}	
where $m_{k}$ and $n_{k}$ are defined in (\ref{Cov{y_p^k,y_p^k}}).
As a result, we obtain that
\begin{align}
	&{\rm{Cov}}\{ \mathbf{q}_k,\mathbf{y}^k_p\} 
	{\rm{Cov}}^{-1}\{ \mathbf{y}^k_p,\mathbf{y}^k_p\} \notag \\
    &=\left(a_{k1} \mathbf{a}_{M}\mathbf{a}_{M}^H \!+\! a_{k2}{\mathbf I}_M\right) \!
    \left\{
     n^{-1}_{k}{\mathbf I}_M \!-\!
    \frac{m_{k}\left(n_{k}\right)^{-2} }{1+Mm_{k}n_{k}^{-1}}\mathbf{a}_{M}\mathbf{a}_{M}^H
    \right\}
    \notag \\
	&=\frac{
		\left\{
			\left(\frac{\sigma^2}{\tau p}\!+\!\varpi\right)a_{k1}\!-\!\varpi\delta a_{k2}
		\right\}
		\mathbf{a}_{M}\mathbf{a}_{M}^H}
		{(a_{k2}\!+\!\frac{\sigma^2}{\tau p}\!+\!\varpi)^2\!+ \! M(a_{k1}\!+\!\varpi\delta)(a_{k2}\!+\!\frac{\sigma^2}{\tau p}\!+\!\varpi)}
		\!+\!\!
		\frac{a_{k2}}{n_{k}}{\mathbf I}_M \!\notag
		\\
	&\triangleq a_{k3}\mathbf{a}_{M}\mathbf{a}_{M}^H + a_{k4}{\mathbf I}_M
	\triangleq{\mathbf A}_k={\mathbf A}_k^H.
\end{align}	

\vspace{0.3cm}
Thus, the LMMSE channel estimate of $\mathbf{q}_k$ is calculated as
\begin{align}
{{\hat{\mathbf q}}_k}&=\mathbb{E}\left\{ \mathbf{q}_k \right\} 
	\!+\!{\rm{Cov}}\{\mathbf{q}_k,\mathbf{y}^k_p\}
	{\rm{Cov}}^{-1}\!\{ \mathbf{y}^k_p,\mathbf{y}^k_p\} 
	(\mathbf{y}_p^k \!-\!\mathbb{E}\left\{ \mathbf{y}_p^k \right\} ) \notag \\
	&={\eta\rho \sqrt {{c_k}\delta {\varepsilon _k}} {{\bf{\overline H}}_2}{\bf{\Phi    }}
		{ \bf{\overline h}}_{k}} + {\mathbf A}_k\left(
		 \mathbf{y}^k_p-\mathbb{E}\left\{ \mathbf{y}^k_p \right\} \right) \notag \\
    &={\mathbf A}_k \mathbf{y}^k_p + \left({\mathbf I}_M- {\mathbf A}_k \right)
	{\eta\rho \sqrt {{c_k}\delta {\varepsilon _k}} {{\bf{\overline H}}_2}{\bf{\Phi    }}
		{ \bf{\overline h}}_{k}} \notag \\
   &\triangleq{\mathbf A}_k \mathbf{y}^k_p+{\mathbf B}_k, 			 
\end{align}	
where
$\mathbf B_k = \left({\mathbf I}_M- {\mathbf A}_k \right)
{\eta\rho \sqrt {{c_k}\delta {\varepsilon _k}} {{\bf{\overline H}}_2}{\bf{\Phi    }}
	{ \bf{\overline h}}_{k}}$.

Then we rewrite the above expression as
\begin{align}
{{\hat{\mathbf q}}_k}&={\mathbf A}_k
\left(
{\mathbf{q}}_k + \frac{\eta}{\sqrt {\tau p}}{\mathbf{H}}_2 \mathbf{\Phi\Theta} \mathbf{V}{\mathbf{s}_{k} }
+\frac{1}{\sqrt {\tau p}}\mathbf{N}{\mathbf{s}_{k}}
\right)
+\mathbf B_k  
\notag \\
&={\mathbf A}_k\left(\sum\limits_{w = 1}^4 {\mathbf{q}_k^w} + \mathbf{d}_k\right)
+{\mathbf A}_k\frac{\eta}{\sqrt {\tau p}}{\mathbf{H}}_2 \mathbf{\Phi\Theta} \mathbf{V}{\mathbf{s}_{k} }\notag \\
&\quad+{\mathbf A}_k\frac{1}{\sqrt {\tau p}}\mathbf{N}{\mathbf{s}_{k}}
\!+\!\left({\mathbf I}_M- {\mathbf A}_k \right)
{\eta\rho \sqrt {{c_k}\delta {\varepsilon _k}} {{\bf{\overline H}}_2}{\bf{\Phi    }}
	{ \bf{\overline h}}_{k}}.
\end{align}	
With the definition of $\sum\limits_{w = 1}^4 {\mathbf{q}_k^w}$, we can get the expression of 
${{\hat{\mathbf q}}_k}$ as (\ref{estiamte q_k}).

Utilizing the estimate ${{\hat {\mathbf q}}_k}$, we obtain the estimation error as
${{\mathbf{e}}_k}={{\mathbf{ q}}_k}-{{\hat{\mathbf q}}_k}$ and the mean of ${{\mathbf{e}}_k}$ is zero. 
As  stated in
\cite{ref35},
we calculate the MSE matrix as
\begin{align}
&{{\mathbf{MSE}}_k}=\mathbb{E}\left\{ \mathbf{e}_k\mathbf{e}_k^H  \right\} \notag \\
&={\rm{Cov}}\{ \mathbf{q}_k,\mathbf{q}_k\}\!-\!{\rm{Cov}}\{\mathbf{q}_k,\mathbf{y}^k_p\}
{\rm{Cov}^{-1}}\!\{\mathbf{y}^k_p,\mathbf{y}^k_p\}
{\rm{Cov}}\{\mathbf{y}^k_p,\mathbf{q}_k\} \notag \\
&=\left({\mathbf I}_M- {\mathbf A}_k \right){\rm{Cov}}\{ \mathbf{q}_k,\mathbf{q}_k\} \notag \\
&\triangleq a_{k5}\mathbf{a}_{M}\mathbf{a}_{M}^H+ a_{k6}{\mathbf I}_M,
\end{align}	
where 
	$a_{k5}\! \!=\! \!\frac
	{a_{k1}\!\left(\frac{\sigma^2}{\tau p}\!+\varpi\right)^2\!
		\!+\! Ma_{k1}\varpi\delta\left(a_{k2}\!+\!\frac{\sigma^2}{\tau p}+ \varpi\right)+ \varpi\delta a^2_{k2}}
	{\left(a_{k2}\!+\!\frac{\sigma^2}{\tau p}+\varpi\right)
		\left(a_{k2}\!+\!\frac{\sigma^2}{\tau p}+\varpi+ M \left(a_{k1}+\varpi\delta\right)\right)
	}$, and 
$a_{k6}=\frac{a_{k2}\left(\frac{\sigma^2}{\tau p} +\varpi   \right)
	}{a_{k2}+\frac{\sigma^2}{\tau p}+\varpi}$. 

\vspace{0.2cm}
Using the ${\rm{\bf MSE}}_k$ matrix, the ${\rm NMSE}_k$ 
is  formulated  as in \cite{ref31}
	\begin{align}
{	\rm NMSE}_k=&\frac{
		{\rm Tr}\left\{ {\rm {{\mathbf {MSE}}}}_k\right\}	}
	{{\rm Tr}
		 \left\{
		{\rm{Cov}}     \left\{
		{\mathbf{q}}_{k},{\mathbf{q}}_{k}
		\right\}
		\right\}
		}
	=\frac{\left(a_{k5}+a_{k6}\right)   }{ \left(a_{k1}+a_{k2}\right)}\label{nmse_k}.
\end{align}

\vspace{0.4cm}
By substituting the values of $m_k$ and $n_k$ into 
(\ref{nmse_k}), we obtain the specific expression of $\rm NMSE_k$.

\vspace{0.5cm}

	\section*{{\large A}PPENDIX {\large D} \\
		{ \large P}ROOF OF {\large T}HEOREM {\large 2}
	}
Before deriving the closed expression, we first give several useful results
used in the following computations.
	Since ${\mathbf A}_k\triangleq a_{k3}\mathbf{a}_{M}\mathbf{a}_{M}^H + a_{k4}{\mathbf I}_M$ and ${ \bf{\overline H}}_{2}=\mathbf{a}_{M}\mathbf{a}_{N}^H $, we can get that
{\begin{align}
&{\rm Tr}\left\{{\mathbf A}_k\right\}=M\left(a_{k3}+a_{k4}\right)\triangleq Me_{k1},\\
&{\mathbf A}_k{ \bf{\overline H}}_{2}=
a_{k3}\mathbf{a}_{M}\mathbf{a}_{M}^H\mathbf{a}_{M}\mathbf{a}_{N}^H+
a_{k4}\mathbf{a}_{M}\mathbf{a}_{N}^H \notag \\ 
&\quad\quad\ \ =\left(Ma_{k3}+a_{k4}\right){ \bf{\overline H}}_{2} 
\triangleq e_{k2}{ \bf{\overline H}}_{2},
\\
&{\rm Tr}\left\{ \!{\mathbf A}_k{\mathbf A}^H_k \!\right\}=\!
{\rm Tr}\left\{\! Ma^2_{k3}\mathbf{a}_{M}\mathbf{a}_{M}^H \!+\!2a_{k3}a_{k4}\mathbf{a}_{M}\mathbf{a}_{M}^H
\!+\! a^2_{k4}\mathbf{I}_{M}  \!    \right\} \notag \\
&\qquad\qquad\quad=M\left( Ma^2_{k3}+2a_{k3}a_{k4}+a^2_{k4}        \right)
\triangleq Me_{k3},
\end{align}
where we define three auxiliary variables $e_{k1}$, $e_{k2}$, and $e_{k3}$. 

\begin{figure*}[ht]
		\setcounter{equation}{86}
	\begin{footnotesize}
		\begin{align}
			&{	E^{ t\, nosie }_k{(\mathbf \Phi)}} =  \frac{\beta}{(\delta+1)}
			\bigg\{			
			\left(M^2N\Delta\delta^2 e^2_{k2}	
			+MN	\Delta\delta e^2_{k2}
			+2M^2\Delta \delta e_{k1}e_{k2}\right)					
			\Big\{  c_k\left[ \varepsilon_k\left(1-\rho^2 \right)+1 \right] +\frac{\varpi}{\Delta}
			\Big \}+M^2\Delta e^2_{k1}c_k(\varepsilon_k+1) \notag \\
			&+\Big\{ 2M^2N\delta a_{k3}a_{k4}+MN(\delta a^2_{k4}+e_{k3})+M^3N\delta a^2_{k3}\Big \}
			\Big\{ \Delta c_k (\varepsilon_k+1)\!+\varpi\!+\!\frac{\sigma^2}{\tau p}+\gamma_k\Big \}
			+|f_k({\mathbf \Phi})|^2\Delta c_k \delta\varepsilon_k\rho^2	
			\left( M^2\delta +  2\frac{M^2}{N}e_{k1}  +M  \right)
			\bigg \}\label{E_phi_t_noise}.
		\end{align}
	\end{footnotesize}
		\hrulefill
	\vspace{0.1cm}
\end{figure*}

\begin{figure*}[hb]
	\hrulefill
	\vspace{0.1cm}
	\begin{footnotesize}
		\begin{align} 
			&	{E^{ nosie }_k(\bf \Phi)}\!=\!{\sigma^2}{E^{ s\, nosie }_k(\bf \Phi)}+
			\eta^2{\sigma^2_e}
			{E^{ t\, nosie }_k(\bf \Phi)}\notag \\
			&	=\sigma^2M\left\{\Delta c_k \delta e_{k2}\left[
			\varepsilon_k\left(1\!-\!\rho^2 \right)+1
			\right]
			+
			\Delta c_k \left(\varepsilon_k \!+\!1 \right)e_{k1}
			+ \gamma_k e_{k1}\right\} 
			+\sigma^2 \frac{M}{N}|f_k({\mathbf \Phi})|^2\Delta  c_k\delta\varepsilon_k\rho^2
			\notag \\
			& +\frac{\sigma_{e}^2\Delta\beta}{N(\delta+1)}
			\bigg\{			
			\left(M^2N\Delta\delta^2 e^2_{k2}	
			+MN	\Delta\delta e^2_{k2}
			+2M^2\Delta \delta e_{k1}e_{k2}\right)					
			\Big\{  c_k\left[ \varepsilon_k\left(1-\rho^2 \right)+1 \right] \!+\frac{\varpi}{\Delta}
			\Big \}\!+M^2\Delta e^2_{k1}c_k(\varepsilon_k+1) 
			\notag \\
			&+\Big\{ 2M^2N\delta a_{k3}a_{k4}\!+\!MN(\delta a^2_{k4}+e_{k3})\!+\!M^3N\delta a^2_{k3}\Big \}
			\Big\{ \Delta c_k (\varepsilon_k+1)\!+\varpi\!+\!\frac{\sigma^2}{\tau p}+\!\gamma_k\Big \}
			\!+\!|f_k({\mathbf \Phi})|^2\Delta c_k \delta\varepsilon_k\rho^2	
			\left( M^2\delta \!+  \!2\frac{M^2}{N}e_{k1} \! +\! M  \right)\bigg \}
			\label{E_phi_noise}
			.
		\end{align}
	\end{footnotesize}
\end{figure*}


Then we focus on deriving each part in (\ref{SINR_k}). Firstly,
we  introduce the derivation of the noise term which is divided into static noise at the BS and thermal noise at the active RIS. For clarity, we define that
%
	\begin{align} 
	\setcounter{equation}{79}
	{E^{ nosie }_k(\bf \Phi)}\!=\!{\sigma^2}{E^{ s\, nosie }_k(\bf \Phi)}+
	\eta^2{\sigma^2_e}
	{E^{ t\, nosie }_k(\bf \Phi)}\label{E_noise}.
		\end{align}
		
Based on the orthogonality property of the LMMSE estimator, the observation vector ${\rm {\bf y}}_p^k$  is orthogonal to estimation error ${{\rm {\bf e}}_k}$, i.e., $\mathbb{E}\left\{ {\rm {\bf y}}_p^k {\bf e}_k\right\}=0$. Since $\mathbb{E}\left\{  {\bf e}_k\right\}=0$, we have
\begin{align} 
	&	\mathbb{E}\left\{ {{\hat{\mathbf q}}^H_k} {{\mathbf{ q}}_k}\right\}=		
		\mathbb{E}\left\{ {{\hat{\mathbf q}}^H_k} \left({\hat{\mathbf{ q}}_k+ {\bf e}_k}\right)
		\right\}\notag \\
	&	=\mathbb{E}\left\{ {{\hat{\mathbf q}}^H_k} {\hat{\mathbf{ q}}_k}\right\}
	\!+\!\mathbb{E}\left\{ {
		\left(	{{\mathbf A}_k \mathbf{y}^k_p\!+\!{\mathbf B}_k}\right)
		^H_k} {{\mathbf{ e}}_k}\right\}
		\!\!=\!\mathbb{E}\left\{ \left \| \hat{\mathbf q}_k\right \|^2 \right\}.
	\end{align}
	
We denote the static noise term in the ${\rm SINR}_k$ as $ \mathbb{E}\left\{ \left \|\hat {\mathbf q}_k\right \|^2 \right\}   
\triangleq E^{ s\,noise }_k(\bf \Phi) $.
Recalling the expressions of (\ref{q_k}) and
(\ref{estiamte q_k}),
 we remove the zero-expectation terms in  $\mathbb{E}\left\{ \left \| \hat{\mathbf q}_k\right \|^2 \right\}$ and derive it as
\begin{align} 
&{E^{ s\, nosie }_k(\bf \Phi)}=
	\mathbb{E}\left\{ \left \| \hat{\mathbf q}_k\right \|^2 \right\}  =
\mathbb{E}\left\{ {{\hat{\mathbf q}}^H_k} {{\mathbf{ q}}_k}\right\}	\notag \\
&=M\left\{\Delta c_k \delta e_{k2}\left[
\varepsilon_k\left(1\!-\!\rho^2 \right)\!+\!1
\right]
 \!+\!
 \Delta c_k \left(\varepsilon_k \!+\!1 \right)e_{k1}\notag
  \!+\! \gamma_k e_{k1}\right\} \\
  &\quad + \frac{M}{N}|f_k({\mathbf \Phi})|^2\Delta  c_k\delta\varepsilon_k\rho^2
  \label{E^s_noise}.
\end{align}

\vspace{0.1cm}
In addition, we define the 
thermal noise term  as ${E^{ t\, nosie }_k(\bf \Phi)}\triangleq
\mathbb{E}\left\{ \left \|\hat {\mathbf q}^H_k
\mathbf{H}_{2} \bf{\Phi\Theta}  
\right \|^2	\right\}$.
By substituting $\mathbf{H}_{2}=\sqrt{\frac{\beta}{\delta+1}} \left (\sqrt {\delta}\, \mathbf{\overline{H}_2}+\tilde{\mathbf{H}}_{2} \right)$ 
into the thermal noise term, 
we can obtain
\begin{align} 
&	\mathbb{E}\left\{ \left \| \hat {\mathbf q}^H_k
	 \mathbf{H}_{2} \bf{\Phi\Theta}  
	\right \|^2	\right\}=\frac{\beta}{\delta+1}\bigg \{
		\mathbb{E}\left\{ \hat{\mathbf q}^H_k\delta{\overline {\mathbf H}}_2 {\overline {\mathbf H}}_2^H\hat{\mathbf q}^H_k	\right\}\notag 
		\\
&		+2\mathbb{E}\left\{\hat{\mathbf q}^H_k\sqrt{\delta} {\overline {\mathbf H}}_2
		 {{{\tilde{\mathbf H}}}^H_2}\hat {\mathbf q}_k
		\right\}
		+\mathbb{E}\left\{\hat{\mathbf q}^H_k 	{{{\tilde{ \mathbf H}}}_2} 
		{{{\tilde{\mathbf H}}}_2}  \hat{\mathbf q}_k
		\right\}
	\bigg \}.
\end{align}

Using Lemma 4 and the independence between ${{{\tilde{\mathbf H}}}_2}$, ${{{\tilde{\mathbf h}}}_k}$, ${{{\tilde{\mathbf d}}}_k}$ and ${\rm{ \mathbf N}}$, 
the specific expression of each term in ${E^{ t\, nosie }_k(\bf \Phi)}$ is calculated respectively as

\begin{align}
	&	\mathbb{E}\left\{ \hat{\mathbf q}^H_k\delta{\overline {\mathbf H}}_2 {\overline {\mathbf H}}_2^H\hat{\mathbf q}^H_k	\right\}
	\notag \\
	&=	M^2\Delta|f_k({\mathbf \Phi})|^2c_k\delta^2\varepsilon_k\rho^2\notag \\
	&	+ \! M^2N\Delta\delta^2e^2_{k2}\left\{
	c_k\left[\varepsilon_k\left(1-\rho^2\right)+1\right] \!+\! \frac{\varpi}{\Delta}	\right\}
	\notag \\
	&+\! M \! N  \delta\left(a_{k3}M \!+ \! a_{k4}\right)^2\!\!
	\left\{
	\Delta c_k \! \left(\varepsilon_k+1\right)\!+\!\varpi\!+\!\frac{\sigma^2}{\tau p}\! +\!\gamma_k
	\right\},
\end{align}

\begin{align}
	&2\mathbb{E}\left\{\hat{\mathbf q}^H_k\sqrt{\delta} {\overline {\mathbf H}}_2
	{{{\tilde{\mathbf H}}}^H_2}\hat {\mathbf q}_k
	\right\}
	\notag \\
	&=2\frac{M^2}{N}|f_k({\mathbf \Phi})|^2\Delta\rho^2 c_k \delta \varepsilon_k e_{k1}
	\notag \\
	&+2M^2\Delta\delta e_{k1}e_{k2} 
	\left\{
	c_k\left[\varepsilon_k\left(1-\rho^2\right)+1\right] \!+\! \frac{\varpi}{\Delta}	\right\},
\end{align}

\begin{align}
	&\mathbb{E}\left\{ \! \hat{\mathbf q}^H_k 	{{{\tilde{ \mathbf H}}}_2} 
	{{{\tilde{\mathbf H}}}_2}  \hat{\mathbf q}_k\!
	\right\}\!
	\notag \\
	& \!=\! \! M |f_k({\mathbf \Phi})|^2\! \Delta c_k \delta \varepsilon_k \rho^2
	\!+\! \! M^2\Delta c_k \! \left(\varepsilon_k \!+\!1\right) \! e_{k1}
	\notag \\
	&+MN \bigg \{ \Delta \delta e^2_{k2}
	\left\{\frac{\varpi}{\Delta}+c_k\left[\varepsilon_k\left(1-\rho^2\right)+1\right]\right\}
	\notag\\
	&\qquad \qquad +e_{k3} \big \{    \gamma_k +\frac{\sigma^2}{\tau p} +\varpi +\Delta c_k \left(\varepsilon_k+1\right)\big \}
	\bigg \}\label{t_noise_3}.
\end{align}

\vspace{0.5cm}
As a  result, we can get the  expression of $E^{ t\, nosie }_k{(\mathbf \Phi)}$
as (\ref{E_phi_t_noise}) at the top of this page.
Finally, we can obtain the noise term  in (\ref{E_noise}).
Based on 
 (\ref{E^s_noise}) and (\ref{E_phi_t_noise}), 
${E^{ nosie }_k(\bf \Phi)}$ is calculated as (\ref{E_phi_noise}) at the bottom of this page.
\begin{figure*}[ht]
	\setcounter{equation}{89}
	\begin{footnotesize}
		\begin{align} 
			{E^{ signal }_k(\bf \Phi)}
			=&M^2\!\left\{\! \Delta c_k \delta e_{k2}\!\left[
			\varepsilon_k\left(1\!-\!\rho^2 \right)\!+\!1
			\right]\!
			\!+\!
			\Delta c_k \! \left(\varepsilon_k \!+\!1 \right) \! e_{k1}\!
			\!+\! \gamma_k e_{k1}\! \right\}^2 
			+ \frac{M^2}{N^2}|f_k({\mathbf \Phi})|^4\Delta^2  c_k^2\delta^2\varepsilon_k^2\rho^4
			\notag \\
			&+
			2\frac{M^2}{N}|f_k({\mathbf \Phi})|^2\Delta  c_k\delta\varepsilon_k\rho^2
			\!\left\{\! \Delta c_k \delta e_{k2}\!\left[
			\varepsilon_k\left(1\!-\!\rho^2 \right)\!+\!1
			\right]\!
			\!+\!
			\Delta c_k \! \left(\varepsilon_k \!+\!1 \right) \! e_{k1}\!
			\!+\! \gamma_k e_{k1}\! \right\}^2
			\label{E_phi_signal}.
		\end{align}
			\hrulefill
			\vspace{0.1cm}
	\end{footnotesize}
\end{figure*}

\vspace{0.3cm}
To the desired signal term, we denote it as ${E^{ signal }_k(\bf \Phi)}$. From the
procedure for obtaining  $E^{ s\,noise }_k(\bf \Phi)$, 
we have known that $	\mathbb{E}\left\{ {{\mathbf{\hat q}}_k} {{\mathbf{ q}}_k}\right\}$ is a real variable, then
we calculate it as 
\begin{align} 
		\setcounter{equation}{88}
	{E^{ signal }_k(\bf \Phi)}=
	|\mathbb{E}\left\{ {{\hat{\mathbf q}}^H_k} {{\mathbf{ q}}_k}\right\}|^2	
	=	\left(	\mathbb{E}\left\{ \left \| \hat{\mathbf q}_k\right \|^2 \right\} \right)^2,
\end{align}
and the specific expression of ${E^{ signal }_k(\bf \Phi)}$ is given as
(\ref{E_phi_signal}) at the top of the next page.


After that, we compute the interference term
$	\mathbb{E} \left\{ |   \hat{\mathbf q}_k^H \mathbf{q}_i| ^2  \right\}   $
	which is denoted as ${I_{ki}{(\bf{\Phi})}}$.  
 Recalling the expressions of  ${{\hat{\mathbf q}}^H_k}$ and ${{{\mathbf q}}_i}$, we decompose ${I_{ki}{(\bf{\Phi})}}$ as
	\begin{align}
			\setcounter{equation}{90}
	&{I_{ki}{(\bf{\Phi})}}=\mathbb{E}\left\{ |{{\hat{\mathbf q}}^H_k} {{\mathbf{ q}}_i}|^2\right\}=\mathbb{E}\bigg\{ 
	\left| {\underline{\hat{{ \mathbf q}}}^H_k}{\underline{\bf{{ q}}}_i}\right|^2	\bigg \}
 +\mathbb{E}\bigg\{ 	\left|{\underline{\hat{{\mathbf q}}}^H_k}{{{\bf d} }_i}\right|^2\bigg \}\notag  \\
&+\mathbb{E}\bigg\{
\left|
\big({\frac{\eta}{\sqrt {\tau p}}\mathbf A_k{\mathbf{H}}_{{2}}}{{\bf{\Phi\Theta} }
	\mathbf{V}
	{\mathbf{s}} }_{ k}\big)^H
{\underline{\bf{{ q}}}_i}
\right|^2\bigg \}\notag \\
&+\mathbb{E}\bigg\{
\left| \big({\frac{\eta}{\sqrt {\tau p}}\mathbf A_k{\mathbf{H}}_{{2}}}{{\bf{\Phi\Theta}} 	
	\mathbf{V}
	{\mathbf{s}} }_{ k}\big)^H
{{{\bf d} }_i} 
\right|^2 \bigg \}\notag \\
&+\mathbb{E}\bigg\{
\left| 
\big(\frac{1}{\sqrt {\tau p}}\mathbf A_k\mathbf{N}{\mathbf{s}_{ k}}\big)^H{\underline{\bf{{ q}}}_i}
 \right|^2\bigg \}
\!+\!\mathbb{E}\bigg\{ \!\!
\left| 
 \big(\frac{1}{\sqrt {\tau p}}\mathbf A_k\mathbf{N}{\mathbf{s}_{ k}}\big)^H \!\!
 {{\bf{{ d}}}_i} \right|^2\!\!\bigg \}
 \notag \\
 &+\mathbb{E}\bigg\{
 \left|  \big( 	{\mathbf A_k}{{\bf{{ d}}}_k}\big)^H{\underline{\bf{{ q}}}_i} \right|^2
 \bigg \}
 +\mathbb{E}\bigg\{\left|   \big( 	{\mathbf A_k}{{\bf{{ d}}}_k}\big)^H{{\bf{{ d}}}_i} \right|^2
	\bigg \}\label{I_KI}.
	\end{align}

\vspace{0.3cm}
Then, we calculate the last seven expectations directly as 
\begin{align}
&	\mathbb{E}\bigg\{\left|   \big( 	{\mathbf A_k}{{\bf{{ d}}}_k}\big)^H{{\bf{{ d}}}_i} \right|^2\bigg \}=\gamma_k\gamma_iMe_{k3}, 
\\
&	\mathbb{E}\bigg\{
	\left|  \big( 	{\mathbf A_k}{{\bf{{ d}}}_k}\big)^H{\underline{\bf{{ q}}}_i} \right|^2
	\bigg \}=
	\frac{M}{N}|f_i({\mathbf \Phi})|^2\Delta c_i \delta\varepsilon_ie^2_{k2}\rho^2\gamma_k
	\notag \\   
&	+\Delta Mc_i\gamma_k\left\{\!
	\delta e^2_{k2}\!\left[\varepsilon_i \! \left(1\!-\!\rho^2\right)\!+\!1\right]\!+\!\left(\varepsilon_i \!+\!1\right)e_{k3}\right\},      
\end{align} 

\begin{align}	
&	\mathbb{E}\bigg\{ \!\!
	\left| 
	\big(\frac{1}{\sqrt {\tau p}}\mathbf A_k\mathbf{N}{\mathbf{s}_{ k}}\big)^H \!\!
	{{\bf{{ d}}}_i} \right|^2\!\!\bigg \}=\frac{\sigma^2}{\tau p}\gamma_i M e_{k3},\\
&	\mathbb{E}\bigg\{
	\left| 
	\big(\frac{1}{\sqrt {\tau p}}\mathbf A_k\mathbf{N}{\mathbf{s}_{ k}}\big)^H{\underline{\bf{{ q}}}_i}
	\right|^2\bigg \} \! =\!
	\frac{M}{N}|f_i({\mathbf \Phi})|^2\Delta c_i \delta\varepsilon_ie^2_{k2}\rho^2
	\frac{\sigma^2}{\tau p}   	\notag \\
	&+
	\Delta Mc_i\frac{\sigma^2}{\tau p}\left\{\!
	\delta e^2_{k2}\!\left[\varepsilon_i \! \left(1\!-\!\rho^2\right)\!+\!1\right]\!+\!\left(\varepsilon_i \!+\!1\right)e_{k3}\right\}, 
\end{align}

\begin{align}	
	&\mathbb{E}\bigg\{\!
	\left| \big({\frac{\eta}{\sqrt {\tau p}}\mathbf A_k{\mathbf{H}}_{{2}}}{{\bf{\Phi\Theta}} 	
		\mathbf{V}
		{\mathbf{s}} }_{ k}\big)^H
	{{{\bf d} }_i} 
	\right|^2\! \bigg \}\!=\!\! M\!\varpi\gamma_i\! \!\left(\delta e^2_{k2}\! +\! e_{k3}\right), 
	\\
&\mathbb{E}\bigg\{\!
\left|
\big({\frac{\eta}{\sqrt {\tau p}}\mathbf A_k{\mathbf{H}}_{{2}}}{{\bf{\Phi\Theta} }
	\mathbf{V}
	{\mathbf{s}} }_{ k}\big)^H\!
{\underline{\bf{{ q}}}_i}
\right|^2\!\bigg \}    \notag \\
&=\Delta\varpi \bigg\{
\frac{M}{N}\!|f_i({\mathbf \Phi})|^2\! c_i\delta\varepsilon_i\rho^2\! e^2_{k2}\notag \\
&+\frac{M}{N}|f_i({\mathbf \Phi})|^2 c_i\delta\varepsilon_i\rho^2e^2_{k2}
\!+\!\!M  c_i \delta^2e^2_{k2}\!\left\{\! \varepsilon_i \! \left(1\!-\!\rho^2\right)\!\!+\!1\!\right\} \notag \\
&+\!\frac{M^2}{N}\!|f_i({\mathbf \Phi})|^2\! c_i\delta^2\!\varepsilon_i\rho^2\! e^2_{k2}\!
+\!\!\frac{2M^2}{N^2}\!|f_i({\mathbf \Phi})|^2\! c_i\delta\varepsilon_i\rho^2\! e_{k1}\! e_{k2}
\notag \\
&+\frac{M^2}{N} c_i \left\{2\delta e_{k1}e_{k2}
\left[\varepsilon_i \! \left(1\!-\!\rho^2\right)\!+\!1\right]
+\left(\varepsilon_i+1\right)e^2_{k1}
\right\} \notag \\	
&+Mc_i \left\{
\left(2\varepsilon_i +2 - \varepsilon_i \rho^2\right)\delta e^2_{k2} + \left(\varepsilon_i+1\right)e_{k3} \right\} \bigg \},
\end{align}
 
\begin{align}	
&\mathbb{E}\bigg\{ 	\left|{\underline{\hat{{\mathbf q}}}^H_k}{{{\bf d} }_i}\right|^2\bigg \}
=
\frac{M}{N}|f_k({\mathbf \Phi})|^2\Delta\gamma_i c_k\delta\varepsilon_k\rho^2
\notag \\
&+M\Delta\gamma_i c_k \left\{\delta e^2_{k2}
\left[\varepsilon_k \! \left(1\!-\!\rho^2\right)\!+\!1\right]
+\left(\varepsilon_k+1\right)e_{k3}
\right\}. 
\end{align}

However, for the first term in (\ref{I_KI}) , we proceed by expanding its expression as
\begin{align}
&\mathbb{E}\bigg\{ 
\left| {\underline{\hat{{ \mathbf q}}}^H_k}{\underline{\mathbf{{ q}}}_i}\right|^2\bigg\}
=\mathbb{E}
\left\{
\left| 
{\sum\limits_{w = 1}^6 ({\hat{\mathbf q}_k^w}})^H	{\sum\limits_{\psi = 1}^4 {\mathbf{ q}_i^\psi}}
\right|^2 
\right\}
\notag \\
&= \sum\limits_{w = 1}^6\sum\limits_{\psi = 1}^4\mathbb{E}\bigg\{
\left| 
{ (  {\hat{\mathbf q}_k^w}} )^H	{ {\mathbf{ q}_i^\psi}}\right|^2 \bigg\} \notag \\
&+\begin{matrix} \sum_{(w_1,\psi_1)\neq(w_2,\psi_2)} 
\! \mathbb{E}\bigg\{ \!\!\left( (
{\hat{\mathbf q}_k^{w_1}} )^H( {\mathbf{ q}_i^{\psi_1}}
)\right)\!\!  
  \left(({\hat{\mathbf q}_k^{w_2}})^H({\mathbf{ q}_i^{\psi_2}})\right)^H \!\!\bigg\}
 \end{matrix}
 \label{ms_cs}.
\end{align}

\vspace{-0.2cm}
Similar to the process described above for calculating other terms, we utilize the independence between random variables and Lemma 4 to compute the modulus-square terms and cross-terms
in (\ref{ms_cs}).

Firstly, we compute one of the 24 modulus-square terms.
When $w = 1$ and $\psi = 1$,  we have
\vspace{0.2cm}
\begin{align}
&\mathbb{E}\bigg\{ 
\left|   \left({\mathbf{{\hat q}}}^1_k\right)^H  {{\bf{{ q}}}^1_i}\right|^2\bigg\}\!=\!
\Delta^2\! c_k\delta^2\! \varepsilon_k c_i \varepsilon_i e^2_{k2}
\!\bigg\{ \!
\frac{M^2}{N^2}\!|{{{\bf{\overline h}}}^H_i}{{{\bf{\overline h}}}_k}|^2\!\!\left(1\!-\!\rho^2\right)^2\notag \\
&+\frac{M^2}{N^2}|F_{ki}|^2\left(l \!-\!\rho^2\right)^2
\!+\!	\frac{M^2}{N}
\left(|f_i({\mathbf \Phi})|^2\!+\!|f_k({\mathbf \Phi})|^2\right)\!
\rho^2\!\left(1\!-\!\rho^2\right)
\notag \\
&+\frac{M^2}{N^2}\!\left(|f_i({\mathbf \Phi})|^2\!+\!|f_k({\mathbf \Phi})|^2\right)\!
2\rho^2\!\left(2\rho^2\!-\!1\!- \! l\right)
\!+ \! M^2\!\!\left(1\!-\!\rho^2\right)^2
\notag \\
&+\frac{M^2}{N}\left(4\rho^2\!+\!4l\rho^2\!- \! l^2\!-\!1\!-\!6\rho^4\right)\!
+\!\frac{M^2}{N^2}|f_i({\mathbf \Phi})|^2|f_i({\mathbf \Phi})|^2\rho^4
\notag \\
&+2\frac{M^2}{N^2}{\rm {Re}}\left\{{{{\bf{\overline h}}}^H_k}{{{\bf{\overline h}}}_i}
f^H_i({\mathbf \Phi})f_k({\mathbf \Phi}) 
\right\}\rho^2\left(1-\rho^2\right) 
\notag \\
&+2\frac{M^2}{N^2}{\rm {Re}}
\left\{  F^H_{ki}
f_k({\mathbf \Phi})f_i({\mathbf \Phi})   \right\}\rho^2\left(l-\rho^2\right)
\bigg\}.
\end{align}
\vspace{0.2cm}
When $w=1$, the remaining terms can be calculated as
\begin{align}
	&\sum\limits_{\psi= 2}^4\mathbb{E}\bigg\{
	\left|   \left({\mathbf{{\hat q}}}^1_k\right)^H  {{\bf{{ q}}}^\psi_i}\right|^2\bigg\}
	=\!\!\eta^4\!c_k\delta\varepsilon_kc_ie_{k2}^2M\!N\!\!\notag \\
	&\qquad\quad\ \ \times \left\{\rho^2|f_k({\mathbf \Phi})|^2+\left(1-\rho^2\right) N \right\}\!
	\left\{\delta M  +\varepsilon_i+1\right\}.
\end{align}
\vspace{0.2cm}
Likewise, when $w \!=\!2,3,4,5,6$, we get other  modulus-square terms  as
\begin{align}
	&\sum\limits_{\psi= 1}^4\!\mathbb{E}\bigg\{\!\!
	\left| \!  \left({\mathbf{{\hat q}}}^2_k\right)^H \!\! {{\bf{{ q}}}^\psi_i}\!\right|^2\!\!\bigg\}
	=\eta^4c_k\delta c_i e^2_{k2}MN^2 
	\left\{\delta M \! + \!\varepsilon_i\! +\!1\right\}\notag \\
&	\!+\!\eta^4c_k\delta c_i\varepsilon_ie_{k2}^2{M^2}N
	\left\{\rho^2|f_i({\mathbf \Phi})|^2\!+\!\left(1\!-\!\rho^2\right)\! N \right\},
\end{align}
\begin{align}
	&\sum\limits_{\psi= 1}^4\!\mathbb{E}\bigg\{\!\!
\left| \!  \left({\mathbf{{\hat q}}}^3_k\right)^H \!\! {{\bf{{ q}}}^\psi_i}\!\right|^2\!\!\bigg\}\notag 
=	\eta^4c_k\varepsilon_kc_i\delta e^2_{k2}MN^2 \\ \notag 
&+\eta^4c_k\varepsilon_kc_i\varepsilon_i\delta e^2_{k2}MN
	\left\{\rho^2|f_i({\mathbf \Phi})|^2\!\!+\!\!\left(\!1\!-\!\rho^2\right)\!\! N\!\right\}\notag\\
&   +\eta^4c_k\varepsilon_kc_i\varepsilon_i
\left\{e^2_{k1}M^2|{{{\bf{\overline h}}}^H_i}{{{\bf{\overline h}}}_k}|^2+e_{k3}MN^2\right\}\notag\\
&+\eta^4c_k\varepsilon_kc_i\left\{
e^2_{k1}M^2N+e_{k3}MN^2
\right\},
\end{align}

\begin{align}
&\sum\limits_{\psi= 1}^4\!\mathbb{E}\bigg\{\!\!
\left| \!  \left({\mathbf{{\hat q}}}^4_k\right)^H \!\! {{\bf{{ q}}}^\psi_i}\!\right|^2\!\!\bigg\}\notag 
=	\eta^4c_kc_i\delta e^2_{k2}MN^2 \\ \notag 
&+\eta^4c_kc_i\delta \varepsilon_ie^2_{k2}MN
\left\{\rho^2|f_i({\mathbf \Phi})|^2\!\!+\!\!\left(\!1\!-\!\rho^2\right)\!\! N\!\right\}\notag\\
&   +\eta^4c_kc_i
\left(\varepsilon_i+1\right)
N
\left\{e^2_{k1}M^2+e_{k3}MN\right\},
\\
&\sum\limits_{\psi= 1}^4\!\mathbb{E}\bigg\{\!\!
\left| \!  \left({\mathbf{{\hat q}}}^5_k\right)^H \!\! {{\bf{{ q}}}^\psi_i}\!\right|^2\!\!\bigg\}\notag \\
&=\eta^4\rho^2c_k\varepsilon_kc_i\varepsilon_i\delta^2M^2|f_k({\mathbf \Phi})|^2
\left\{\rho^2|f_i({\mathbf \Phi})|^2\!\!+\!\!\left(1\!-\!\rho^2\right)\!N\right\} 
\notag \\
&\quad+\eta^4\rho^2c_k\delta\varepsilon_kc_iMN|f_k({\mathbf \Phi})|^2
\left\{\delta M \! + \!\varepsilon_i\! +\!1\right\},
\end{align}
\begin{align}
&\sum\limits_{\psi= 1}^4\!\mathbb{E}\bigg\{\!\!
\left| \!  \left({\mathbf{{\hat q}}}^6_k\right)^H \!\! {{\bf{{ q}}}^\psi_i}\!\right|^2\!\!\bigg\} \notag \\
&=
\sum\limits_{\psi= 1}^4\!\mathbb{E}\bigg\{\!
\left| \!  \left(
-\eta\rho \sqrt {{c_k}\delta {\varepsilon _k}}   \mathbf A_k            
 {{{\bf{\overline H}}}_2}{{\bf{\Phi }}{\overline {\mathbf h}_k}}
\right)^H \!\!\! {{\bf{{ q}}}^\psi_i}\!\right|^2\!\bigg\}\notag \\
&=\eta^4e^2_{k2}\rho^2c_k\varepsilon_kc_i\varepsilon_i\delta^2M^2|f_k({\mathbf \Phi})|^2
\left\{\rho^2|f_i({\mathbf \Phi})|^2\!\!+\!\!\left(1\!-\!\rho^2\right)\!N\right\} 
\notag \\
&\quad+\eta^4e^2_{k2}\rho^2c_k\delta\varepsilon_kc_iMN|f_k({\mathbf \Phi})|^2
\left\{\delta M \! + \!\varepsilon_i\! +\!1\right\},
\end{align}




Additionally, although there is a total of 40 non-zero terms in the cross-terms, half of them are complex conjugates of the other half. Hence, we only need to calculate 20 non-zero terms.
As a result, we have
\begin{align}
	\begin{matrix} 
		\sum_{(w_1,\psi_1)\neq(w_2,\psi_2)} 
		\! \mathbb{E}\bigg\{ \!\!\left( (
		{\hat{\mathbf q}_k^{w_1}} )^H( {\mathbf{ q}_i^{\psi_1}}
		)\right)\!\!  
		\left(({\hat{\mathbf q}_k^{w_2}})^H({\mathbf{ q}_i^{\psi_2}})\right)^H \!\!\bigg\}\\
\!	\!=\!2{\rm Re}\!\left\{\!\sum_{w_1,\psi_1,w_2,\psi_2} 
	\! \mathbb{E}\bigg\{ \!\!\!\left( (
	{\hat{\mathbf q}_k^{w_1}} )^H( {\mathbf{ q}_i^{\psi_1}}
	)\right)\!\!  
	\left(({\hat{\mathbf q}_k^{w_2}})^H({\mathbf{ q}_i^{\psi_2}})\right)^H \!\!\bigg\}\!\!\right\}\label{2re_cross_term}.
	\end{matrix}
\end{align}

Herein, we use $ {{\rm CS}_{ki}}$ to represent the sum of all these cross-terms. Then, we calculate these  non-zero terms one by one 
 in 
(\ref{2re_cross_term})
and we get 
(\ref{CS}) at the bottom of this page.
\begin{figure*}[hb]
		\hrulefill
		\vspace{0.2cm}
			\begin{footnotesize}
	\begin{align}
		&{\rm CS}_{ki}
		\!=\!2\eta^4c_k\varepsilon_k\delta c_i \varepsilon_i e_{k1} e_{k2}M^2
		{\rm Re}\Bigl\{
		\rho^2 f_i({\mathbf \Phi})f^H_k({\mathbf \Phi})  
		{{{\bf{\overline h}}}^H_i}{{{\bf{\overline h}}}_k}
		\!\!	+\!\left(1\!\!-\!\rho^2\right)\!
		|{{{\bf{\overline h}}}^H_k}{{{\bf{\overline h}}}_i}|^2\!
		\Bigr\}
		+2\eta^4c_k\delta c_i e_{k1}e_{k2}M^2N
		+2\eta^4\rho^2c_k\delta\varepsilon_kc_ie_{k1}M^2| f_k({\mathbf \Phi})  |^2\left(1-e_{k2}\right)
		\notag \\
		&+\!2\eta^4{M^2}{\rm Re}\Bigl\{ {{{\bf{\overline h}}}^H_k}{{{\bf{\overline h}}}_i}
		f^H_i({\mathbf \Phi})f_k({\mathbf \Phi})
		\Bigr\}\!\!
		 \left\{
		\rho^2\left(1-\rho^2\right)c_k\delta^2\varepsilon_kc_i\varepsilon_ie_{k2}\!
		\left(1\! -\! e_{k2}\right)
		\right\}
		\!+\!2\eta^4{M^2}{\rm Re}\Bigl\{ F^H_{ki}f_i({\mathbf \Phi})f_k({\mathbf \Phi})
		\Bigr\}
		\left\{
	\rho^2\left(l \!-\!\rho^2\right)c_k\delta^2\varepsilon_kc_i\varepsilon_ie_{k2}\!
		\left(1\! -\! e_{k2}\right)
		\right\}
		\notag \\
		&+2\eta^4{M^2}| f_k({\mathbf \Phi})  |^2\rho^2c_k\delta^2\varepsilon_kc_i
		\varepsilon_ie_{k2}\left(1-e_{k2}\right)\left(2\rho^2-l-1\right)
		+
		2\eta^4{M^2}N| f_k({\mathbf \Phi})  |^2\Delta^2\rho^2\left(1-\rho^2\right)
		c_k\delta^2\varepsilon_kc_i\varepsilon_ie_{k2}\left(1-e_{k2}\right)
		\notag \\
	&+2\eta^4M^2| f_k({\mathbf \Phi})  |^2\!| f_i({\mathbf \Phi})  |^2\!
	\rho^4\!c_k\delta^2 \! \varepsilon_kc_i\varepsilon_ie_{k2}\!\left(1\!-\!e_{k2}\right)\!
	\!+\!2\eta^4\!c_k\delta\varepsilon_kc_ie_{k1}\! e_{k2}M^2\!\left\{
	\!\rho^2\!| f_k({\mathbf \Phi})  |^2 \!\!+\!\! \left(1\!-\!\rho^2\right)\!\!N\!
	\right\}
\!+\!2\eta^4\!c_k\delta c_i\varepsilon_ie_{k1}\! e_{k2}M^2\!\left\{
\!\rho^2| f_i({\mathbf \Phi})  |^2 \!\!+\!\! \left(1\!-\!\rho^2\right)\!\!N\!
\right\}\notag \\
%
&+2\eta^4\!\rho^2c_k\delta\varepsilon_kc_ie_{k2}M\!N| f_k({\mathbf \Phi})  |^2\!\!\left(1\!-\!e_{k2}\right)\left(\delta M + \varepsilon_i +1\right)
+2\eta^4\rho^2c_k\delta\varepsilon_kc_i\varepsilon_ie_{k1}M^2
{\rm Re}\Bigl\{
{{{\bf{\overline h}}}^H_k}{{{\bf{\overline h}}}_i}
f^H_i({\mathbf \Phi})f_k({\mathbf \Phi})  
\Bigr\}\left(1-e_{k2}\right)
\notag \\
%
&-2\eta^4\rho^2c_k\delta^2\varepsilon_kc_i\varepsilon_ie_{k2}M^2| f_k({\mathbf \Phi})  |^2
\left\{
\rho^2| f_i({\mathbf \Phi})  |^2+\left(1-\rho^2\right)N
\right\}
-2\eta^4\rho^2c_k\delta\varepsilon_kc_ie_{k2}MN| f_k({\mathbf \Phi})  |^2\left(\delta M+\varepsilon_i+1\right)\label{CS}.
	\end{align} 
		\end{footnotesize}
\end{figure*}
\begin{figure*}[ht]
	\begin{footnotesize}
		\begin{align}
			&{{I }_{ki}(\mathbf \Phi)}=
			\frac{M^2}{N^2}|{{{\bf{\overline h}}}_{{k}}^H}{{{\bf{\overline h}}}_{{i}}}|^2
			\Delta^2{c_k}\varepsilon_k{c_i}\varepsilon_i
			\left[\delta e_{k2}(1-\rho^2) +  e_{k1}\right]^2
			+\frac{M^2}{N^2}|F_{ki}|^2\Delta^2{c_k}\delta^2\varepsilon_k{c_i}\varepsilon_ie_{k2}^2
			\left(l-\rho^2\right)^2 
			+\frac{M^2}{N^2}| f_k({\mathbf \Phi})  |^2| f_i({\mathbf \Phi})|^2
			\Delta^2\rho^4{c_k}\delta^2\varepsilon_k{c_i}\varepsilon_i \notag \\
			+&2\frac{M^2}{N^2}\Delta^2\rho^2c_{i}\delta
			\bigg\{
			c_{k}\delta\varepsilon_{k}\varepsilon_{i}e_{k2}(2\rho^2-1-l)
			\Big\{ | f_k({\mathbf \Phi})  |^2+| f_i({\mathbf \Phi})  |^2e_{k2}	\Big \}
			+| f_k({\mathbf \Phi})  |^2c_{k}\varepsilon_{k}e_{k1}
			+| f_i({\mathbf \Phi})  |^2\varepsilon_{i}e_{k1}e_{k2}
			\left(c_k+\frac{\varpi}{\Delta}\right)
			\bigg \} \notag \\
			+&\frac{M^2}{N}\Delta^2\rho^2c_{i}\delta^2
			\bigg\{c_{k}\varepsilon_{k}\varepsilon_{i}(1-\rho^2)
			\Big\{ | f_k({\mathbf \Phi})  |^2+| f_i({\mathbf \Phi})  |^2e^2_{k2}	\Big \}
			+| f_k({\mathbf \Phi})  |^2c_{k}\varepsilon_{k}
			+| f_i({\mathbf \Phi})  |^2\varepsilon_{i}e_{k2}^2
			\left(c_k+\frac{\varpi}{\Delta}\right)
			\bigg \}        \notag \\
			+&\frac{M}{N}\Delta\rho^2\delta
			\bigg\{ \Delta c_{k}\varepsilon_{k}c_{i}\varepsilon_{i}
			\Big\{ | f_k({\mathbf \Phi})  |^2+| f_i({\mathbf \Phi})  |^2e^2_{k2}	\Big \}
			+c_{k}\varepsilon_k| f_k({\mathbf \Phi})  |^2(\Delta c_{i} +\gamma_i)
			+| f_i({\mathbf \Phi})|^2c_{i}\varepsilon_{i}e^2_{k2}
			\left(\frac{\sigma^2}{\tau p}+\varpi+\gamma_k+\Delta c_k\right)
			\bigg \}  \notag \\
			+&M^2\Delta^2{c_i}\delta^2e_{k2}^2\bigg\{ 
			{c_k}(1-\rho^2)\Big\{ \varepsilon_{i}+\varepsilon_{k}+\varepsilon_{i}\varepsilon_{k}(1-\rho^2)\Big \}+c_k
			+\frac{\varpi}{\Delta}
			\Big\{   1+ \varepsilon_{i} (1-\rho^2)\Big \}
			\bigg \} 
			+\frac{M^2}{N}\Delta^2{c_i}
			e_{k1}\left({c_k}+\frac{\varpi}{\Delta}\right)(2\delta e_{k2}+e_{k1})
			\notag \\ 
			+&\frac{M^2}{N}\Delta^2{c_i}\bigg\{ 
			{c_k}\varepsilon_k\delta^2\varepsilon_ie_{k2}^2(4\rho^2+4l\rho^2-l^2-1-6\rho^4)
			+e_{k1}\Big\{
			\varepsilon_i\left({c_k}+\frac{\varpi}{\Delta}\right)+c_k\varepsilon_{k}
			\Big \}\Big\{ 2\delta e_{k2}(1-\rho^2)+e_{k1} \Big \}
			\bigg \} \notag \\ 
			+&M\bigg\{ 
			\Delta^2c_{k}c_{i}\Big\{ 
			(\varepsilon_k\varepsilon_i+\varepsilon_k+\varepsilon_i+1)(2\delta e_{k2}^2+e_{k3})
			\!-\!\rho^2\delta e_{k2}^2(2\varepsilon_k\varepsilon_i+\varepsilon_k+\varepsilon_i)
			\Big \}\!
			\!+ \! \varpi
			\Big\{(\delta e_{k2}^2+e_{k3})\gamma_i \!+\!
			\Delta c_i(\varepsilon_i\!+\!1)(2\delta e_{k2}^2\!+\!e_{k3})\!-\!\Delta\delta e_{k2}^2 c_i \varepsilon_i\rho^2 \Big \}\notag \\
			+&\left(\frac{\sigma^2}{\tau p}\!+\!\gamma_k\right)
			\Big\{ \Delta c_i(\varepsilon_i+1)(\delta e_{k2}^2+e_{k3})
			-\Delta \delta e_{k2}^2c_i \varepsilon_i\rho^2+e_{k3}\gamma_i\Big \}
			+\Delta c_k \gamma_i(\varepsilon_k+1)(\delta e_{k2}^2+e_{k3})
			-\Delta\delta e_{k2}^2c_k\gamma_i\varepsilon_k\rho^2
			\bigg \} \notag \\
			+&2\frac{M^2}{N^2}{\rm {Re}}
			\left\{  {{{\bf{\overline h}}}^H_k}{{{\bf{\overline h}}}_i}
			f^H_i({\mathbf \Phi})f_k({\mathbf \Phi})   \right\}\Delta^2\rho^2c_k\delta\varepsilon_k
			c_i\varepsilon_i\left\{\delta e_{k2} \left(1-\rho^2\right)+e_{k1}\right\}
			+2\frac{M^2}{N^2}{\rm {Re}}
			\left\{  F^H_{ki}
			f_k({\mathbf \Phi})f_i({\mathbf \Phi})   \right\}\Delta^2\rho^2
			\left(l-\rho^2\right)
			c_k\delta^2\varepsilon_k
			c_i\varepsilon_i e_{k2}\label{I_phi_ki}.
		\end{align}
	\end{footnotesize}
		\hrulefill
	\vspace{0.2cm}
\end{figure*}

Finally, we can complete
the derivation of $ {I_{ki}{(\bf{\Phi})}}$ 
with some simplifications. We express it as (\ref{I_phi_ki}) which is 
shown at the top of the next page. 
In addition, during the calculation process, 
we define some phase shifts related variables as below
\begin{align}
	&f_k({\mathbf \Phi})\triangleq\mathbf{a}^H_N   {\bf{\Phi}}{{{\bf{\overline h}}}_k}
		\triangleq{\sum\limits_{n = 1}^N}  f_{k,n}   ({\mathbf \Phi})
		={\sum\limits_{n = 1}^N}e^{j  (  \xi _{\mathrm{n}}^{k}+ \theta_n)  },\\
		&\xi _{\mathrm{n}}^{k}=2\pi\frac{d}{\lambda}
		\left(   
		{ \lfloor \left( n-1 \right) /\sqrt{N} \rfloor  } \notag
		\left(  {\sin \varphi  _{\mathrm{kr}}^{a}\sin \varphi  _{\mathrm{kr}}^{e}-\sin \varphi _{\mathrm{t}}^{a}\sin \varphi _{\mathrm{t}}^{e}}                    \right) 
		\right.\\& \left. 
		\quad+{\left(      \left( n-1 \right) \mathrm{mod}\sqrt{N}\right) }
		{\left( \cos\varphi   _{\mathrm{kr}}^{e}-\cos \varphi _{\mathrm{t}}^{e} \right) }     
		\right), \\
		&F_{ki}\triangleq\sum\limits_{n = 1}^N f_{k,n}   ({\mathbf \Phi}) f_{i,n}   ({\mathbf \Phi}).
\end{align}

Following that, we calculate the signal leakage term which can be expressed as
\begin{align}
	 E^{ leak }_k({\bf \Phi})=	 
\mathbb{E}  \left\{ | {\hat{\mathbf q}}_k^H {\mathbf{q}}_k|^2  \right\}
-
| \mathbb{E} \left\{ \hat{\mathbf q}_k^H \mathbf{q}_k\right\}|^2,
\end{align}
where $ \mathbb{E} \left\{\hat{ \mathbf q}_k^H \mathbf{q}_k\right\} $ is known as (\ref{E^s_noise}).
Therefore, we only need to
derive the expectation of $\mathbb{E}  \left\{ | {\hat{\mathbf q}}_k^H {\mathbf{q}}_k|^2  \right\} $.
Similar to the calculation of ${{I }_{ki}(\bf \Phi)}$, we eliminate terms with zero expectation
by utilizing the independence  between the direct channel, the cascaded channel, and two kinds of
noise.
Finally, 
$\mathbb{E}  \left\{ | \hat {\mathbf q}_k^H \mathbf{q}_k|^2  \right\}$ can be expanded as

	\begin{align}
	&\mathbb{E}  \left\{ | \hat{\mathbf q}_k^H \mathbf{q}_k|^2  \right\}
	=\mathbb{E}\bigg\{ 
	\left| {\underline{\hat {{\mathbf q}}}^H_k}{\underline{\mathbf{{ q}}}_k}\right|^2\bigg \}
	+\mathbb{E}\bigg\{ 	\left|{\underline{\hat{{\mathbf q}}}^H_k}{{{\mathbf d} }_k}\right|^2\bigg \}\notag  \\
	&+ \mathbb{E}\bigg\{  \left|
	\big({\frac{\eta}{\sqrt {\tau p}}\mathbf A_k{\bf{H}}_{{2}}}{{\bf{\Phi\Theta}} 	
		\mathbf{V}
		{\mathbf{s}} }_{ k}\big)^H
	{\underline{\bf{{ q}}}_k}
	\right|^2\bigg \} \notag  \\
	&+\mathbb{E}\bigg\{\left| \big({\frac{\eta}{\sqrt {\tau p}}\mathbf A_k{\bf{H}}_{{2}}}{{\bf{\Phi\Theta} }
	\mathbf{V}
		{\mathbf{s}} }_{ k}\big)^H
	{{{\bf d} }_k} 
	\right|^2\bigg \} \notag \\
	&+ \mathbb{E}\bigg\{ \left| 
	\big(\frac{1}{\sqrt {\tau p}}\mathbf A_k\mathbf{N}{\mathbf{s}_{ k}}\big)^H{\underline{\bf{{ q}}}_k}
	\right|^2\bigg \}
	\!+\! \mathbb{E}\bigg\{ \!\!\left| 
	\big(\frac{1}{\sqrt {\tau p}}\mathbf A_k\mathbf{N}{\mathbf{s}_{ k}}\big)^H \!\!{{\bf{{ d}}}_k} \right|^2\!\!\bigg \}
	\notag \\
	&+ \mathbb{E}\bigg\{ \left|  \big( 	{\mathbf A_k}{{\bf{{ d}}}_k}\big)^H{\underline{\bf{{ q}}}_k} \right|^2\bigg \}
	+\mathbb{E}\bigg\{ \left|   \big( 	{\mathbf A_k}{{\bf{{ d}}}_k}\big)^H{{\bf{{ d}}}_k} \right|^2\bigg \}
	\notag \\
	&+2{\rm Re} \bigg \{ \mathbb{E}\bigg\{ \left|  
	 {\underline{\hat{{\mathbf q}}}^H_k}{\underline{\bf{{ q}}}_k}{{\bf{{ d}}}^H_k}
	 \mathbf A_k{{\bf{{ d}}}_k}
\right	|^2   \bigg\}	\bigg \}.
\end{align}
                 
For the calculation of $\mathbb{E}  \left\{ | \hat{\mathbf q}_k^H \mathbf{q}_k|^2  \right\}$,
we omit the computation process of the last eight terms as they are straightforward and simple.
Particularly, our main task is to compute the first term among the nine expectations. 
For the first term, we also expand it as
\begin{align}
	&\mathbb{E}\bigg\{ 
	\left| {\underline{\hat{{\mathbf q}}}^H_k}{\underline{\mathbf{{ q}}}_k}\right|^2\bigg\}
	=\mathbb{E}
	\left\{
	\left| 
	{\sum\limits_{w = 1}^6 ({\hat{\mathbf q}_k^w}})^H	{\sum\limits_{\psi = 1}^4 {\mathbf{ q}_k^\psi}}
	\right|^2 
	\right\}
	\notag \\
	&= \sum\limits_{w = 1}^6\sum\limits_{\psi = 1}^4\mathbb{E}\bigg\{
	\left| 
	{ (  {\hat{\mathbf q}_k^w}} )^H	{ {\mathbf{ q}_k^\psi}}\right|^2 \bigg\} \notag \\
	&+\begin{matrix} \sum_{(w_1,\psi_1)\neq(w_2,\psi_2)} 
		\mathbb{E}\bigg\{ \!\left( \! (
		{\hat{\mathbf q}_k^{w_1}} )^H( {\mathbf{ q}_k^{\psi_1}}
		)\right) \!\! 
		\left(({\hat{\mathbf q}_k^{w_2}})^H({\mathbf{ q}_k^{\psi_2}})\right)^H \!\!\bigg\}
	\end{matrix}.
\end{align}

There are 24 modulus-square terms and 64 non-zeros cross-terms. 
Similarly, half of cross-terms are complex conjugates of the other half.
As the derivation process of the interference term, we calculate the expectations of non-zero terms.
Finally, the expression of  $E^{ leak }_k(\bf \Phi)$ is presented  as (\ref{E_phi_leak}) at the top of the next page.
\begin{figure*}[ht]
	\begin{footnotesize}
		\begin{align}
			&E_k^{leak}{(\bf{\Phi})}  =
			M^2\Delta^2\bigg\{
			c_{k}\delta^2e^2_{k2}\left[\varepsilon_k(1\!-\!\rho^2) \!+\!1\right]
			\Big\{c_k \left[\varepsilon_k(1\!-\!\rho^2) \!+\!1\right]+  \frac{\varpi}{\Delta} \Big \}
			+2 c^2_k\delta e_{k1}e_{k2} (e_{k2}\!-\!1)
			\bigg \} 
			+2\frac{M}{N^2}\left|f_k({\mathbf \Phi})\right|^2\Delta^2\rho^2c^2_k\delta\varepsilon_k
			e_{k2}(1\!+\! e_{k2})
			\notag \\
			&+\frac{M^2}{N}\Delta^2 c_k\bigg\{
			e_{k1}\Big\{
			\varepsilon_k(2c_{k}+\frac{\varpi}{\Delta})\left[2\delta e_{k2}(1-\rho^2)+e_{k1}\right]+e_{k1}(c_k+\frac{\varpi}{\Delta})+2\delta e_{k2}
			(c_ke_{k2}+\frac{\varpi}{\Delta})
			\Big\}+
			c_{k}\delta^2e^2_{k2}\left[\varepsilon_k^2
			(4l\rho^2\!+\!4\rho^2\!-\!l^2\!-\!6\rho^4\!-\!1)\! - \! l^2\right]\!
			\bigg \} \notag \\
			&+\varpi\delta e^2_{k2}\Big\{ \Delta c_{k}(2\varepsilon_k+2\!-\!\varepsilon_k\rho^2)\!+\!\gamma_k
			\Big \}
			\!	+\!\gamma_k e_{k3}(\frac{\sigma^2}{\tau p}\!+\!\gamma_k \!+\!\varpi)
			\bigg \}
			\!+\!2\frac{M^2}{N^2}\!\left|f_k({\mathbf \Phi})\right|^2\!\Delta^2\!{c_k}\delta\varepsilon_k\rho^2\!
			\bigg\{c_k(1+e_{k2})\left[ \delta\varepsilon_k(2\rho^2-1-l)e_{k2} +e_{k1}  \right]\!+\!
			\frac{\varpi}{\Delta}e_{k1}e_{k2}
			\bigg \}  
			\notag \\
			&+\frac{M^2}{N^2}\Delta^2c^2_k\delta^2\varepsilon_k^2e_{k2}(l-\rho^2)
			\bigg\{
			\left|F_{kk}({\mathbf \Phi})\right|^2e_{k2}(l-\rho^2)
			+2{\rm{Re}}\Big\{ f_k({\mathbf \Phi})^2 F_{kk}^H \Big\}\rho^2
			\bigg \}
			+\frac{M^2}{N^2}\Delta^2{c_k^2}\delta^2e_{k2}^2l^2\left|
			\mathbf{a}_N^T{\bf{\Phi }}^H{\bf{\Phi }}^H\mathbf{a}_N 
			\right|^2
			\notag \\
			&+\frac{M}{N}\left|f_k({\mathbf \Phi})\right|^2\Delta c_k\delta\varepsilon_k\rho^2
			\bigg\{ \Delta c_k(\varepsilon_k+1)(e_{k2}+1)+\gamma_k+(\varpi+
			\frac{\sigma^2}{\tau p}+\gamma_k)e^2_{k2}
			\bigg \} 
			+\frac{M}{N}\Delta^2c^2_{k}\bigg\{
			(2\varepsilon_k+1)(2\delta e^2_{k2}+e_{k3})-4\delta\varepsilon_k\rho^2e^2_{k2}
			\bigg \} \notag \\
			&+M\bigg\{ \Delta c_{k}	\Big\{ \Delta c_{k}	(\varepsilon_k+1)+2\gamma_k+\frac{\sigma^2}{\tau p}\Big \}
			\Big\{ (\delta e^2_{k2}+e_{k3})(\varepsilon_k+1)-\delta\varepsilon_k\rho^2e^2_{k2}\Big \}
			+ \Delta c_{k}(\varepsilon_k+1)\Big\{
			\Delta c_{k}\delta e^2_{k2}\left[\varepsilon_k(1-\rho^2)+1\right]+\varpi e_{k3}\Big \}
			\notag \\
			&+\frac{M^2}{N}\left|f_k({\mathbf \Phi})\right|^2\Delta^2c_k\delta^2\varepsilon_k\rho^2
			\bigg\{ c_k
			\Big\{ (e^2_{k2}-1)^2 + 2e_{k1} + \varepsilon_k(1-\rho^2)(e^2_{k2}+1)
			\Big\}
			+\frac{\varpi}{\Delta}e^2_{k2}
			\bigg \}\label{E_phi_leak}. 
		\end{align}
	\end{footnotesize}
	\vspace{-0.65cm}
	\hrulefill
\end{figure*}

\bibliographystyle{IEEEtran}
\bibliography{IEEEabrv,Refer}

\begin{thebibliography}{10}
\providecommand{\url}[1]{#1}
\csname url@samestyle\endcsname
\providecommand{\newblock}{\relax}
\providecommand{\bibinfo}[2]{#2}
\providecommand{\BIBentrySTDinterwordspacing}{\spaceskip=0pt\relax}
\providecommand{\BIBentryALTinterwordstretchfactor}{4}
\providecommand{\BIBentryALTinterwordspacing}{\spaceskip=\fontdimen2\font plus
\BIBentryALTinterwordstretchfactor\fontdimen3\font minus
  \fontdimen4\font\relax}
\providecommand{\BIBforeignlanguage}[2]{{%
\expandafter\ifx\csname l@#1\endcsname\relax
\typeout{** WARNING: IEEEtran.bst: No hyphenation pattern has been}%
\typeout{** loaded for the language `#1'. Using the pattern for}%
\typeout{** the default language instead.}%
\else
\language=\csname l@#1\endcsname
\fi
#2}}
\providecommand{\BIBdecl}{\relax}
\BIBdecl

\bibitem{ref1}
Q.~Wu and R.~Zhang, ``Intelligent reflecting surface enhanced wireless network
  via joint active and passive beamforming,'' \emph{{IEEE} Trans. Wireless
  Commun.}, vol.~18, no.~11, pp. 5394--5409, Nov. 2019.

\bibitem{ref2}
Q.~Tao, J.~Wang, and C.~Zhong, ``Performance analysis of intelligent reflecting
  surface aided communication systems,'' \emph{{IEEE} Commun. Lett.}, vol.~24,
  no.~11, pp. 2464--2468, Nov. 2020.

\bibitem{ref3}
C.~Pan et~al., ``Multicell {MIMO} communications relying on intelligent
  reflecting surfaces,'' \emph{{IEEE} Trans. Wireless Commun.}, vol.~19, no.~8,
  pp. 5218--5233, Aug. 2020.

\bibitem{ref4}
C.~Huang et~al., ``Holographic {MIMO} surfaces for {6G} wireless networks:
  Opportunities, challenges, and trends,'' \emph{{IEEE} Wireless Commun.},
  vol.~27, no.~5, pp. 118--125, Oct. 2020.

\bibitem{sanguinetti2020toward}
L.~Sanguinetti, E.~Bj{\"o}rnson, and J.~Hoydis, ``Toward massive {MIMO} 2.0:
  Understanding spatial correlation, interference suppression, and pilot
  contamination,'' \emph{{IEEE} Trans. Commun.}, vol.~68, no.~1, pp. 232--257,
  Jan. 2020.

\bibitem{chen2021analysis}
Z.~Peng, X.~Chen, W.~Xu, C.~Pan, L.-C. Wang, and L.~Hanzo, ``Analysis and
  optimization of massive access to the {IoT} relying on multi-pair two-way
  massive {MIMO} relay systems,'' \emph{{IEEE} Trans. Commun.}, vol.~69, no.~7,
  pp. 4585--4598, Jul. 2021.

\bibitem{nguyen2022uav}
M.-H.~T. Nguyen, E.~Garcia-Palacios, T.~Do-Duy, O.~A. Dobre, and T.~Q. Duong,
  ``{UAV}-aided aerial reconfigurable intelligent surface communications with
  massive {MIMO} system,'' \emph{IEEE Trans. Cogn. Commun. Netw.}, vol.~8,
  no.~4, pp. 1828--1838, Dec. 2022.

\bibitem{wang2021joint}
P.~Wang, J.~Fang, L.~Dai, and H.~Li, ``Joint transceiver and large intelligent
  surface design for massive {MIMO} mm{W}ave systems,'' \emph{{IEEE} Trans.
  Wireless Commun.}, vol.~20, no.~2, pp. 1052--1064, Feb. 2021.

\bibitem{he2022reconfigurable}
J.~He, K.~Yu, Y.~Shi, Y.~Zhou, W.~Chen, and K.~B. Letaief, ``Reconfigurable
  intelligent surface assisted massive {MIMO} with antenna selection,''
  \emph{{IEEE} Trans. Wireless Commun.}, vol.~21, no.~7, pp. 4769--4783, Jul.
  2022.

\bibitem{bjornson2020intelligent}
E.~Bj{\"o}rnson, {\"O}.~{\"O}zdogan, and E.~G. Larsson, ``Intelligent
  reflecting surface versus decode-and-forward: How large surfaces are needed
  to beat relaying?'' \emph{{IEEE} Wireless Commun. Lett.}, vol.~9, no.~2, pp.
  244--248, Feb. 2020.

\bibitem{ref15}
Y.~Han, W.~Tang, S.~Jin, C.-K. Wen, and X.~Ma, ``Large intelligent
  surface-assisted wireless communication exploiting statistical {CSI},''
  \emph{{IEEE} Trans. Veh. Technol.}, vol.~68, no.~8, pp. 8238--8242, Aug.
  2019.

\bibitem{ref16}
M.-M. Zhao, Q.~Wu, M.-J. Zhao, and R.~Zhang, ``Intelligent reflecting surface
  enhanced wireless networks: Two-timescale beamforming optimization,''
  \emph{{IEEE} Trans. Wireless Commun.}, vol.~20, no.~1, pp. 2--17, Jan. 2021.

\bibitem{ref17}
A.~Abrardo, D.~Dardari, and M.~Di~Renzo, ``Intelligent reflecting surfaces:
  Sum-rate optimization based on statistical position information,''
  \emph{{IEEE} Trans. Commun.}, vol.~69, no.~10, pp. 7121--7136, Oct. 2021.

\bibitem{ref13}
C.~Hu, L.~Dai, S.~Han, and X.~Wang, ``Two-timescale channel estimation for
  reconfigurable intelligent surface aided wireless communications,''
  \emph{{IEEE} Trans. Commun.}, vol.~69, no.~11, pp. 7736--7747, Nov. 2021.

\bibitem{2022zhikangdapower}
K.~Zhi, C.~Pan, H.~Ren, and K.~Wang, ``Power scaling law analysis and phase
  shift optimization of {RIS}-aided massive {MIMO} systems with statistical
  {CSI},'' \emph{{IEEE} Trans. Commun.}, vol.~70, no.~5, pp. 3558--3574, May.
  2022.

\bibitem{zhi2022two-timescale}
K.~Zhi et~al., ``Two-timescale design for reconfigurable intelligent
  surface-aided massive {MIMO} systems with imperfect {CSI},'' \emph{{IEEE}
  Trans. Inf. Theory}, vol.~69, no.~5, pp. 3001--3033, May. 2023.

\bibitem{ref5}
M.~Najafi, V.~Jamali, R.~Schober, and H.~V. Poor, ``Physics-based modeling and
  scalable optimization of large intelligent reflecting surfaces,''
  \emph{{IEEE} Trans. Commun.}, vol.~69, no.~4, pp. 2673--2691, Apr. 2021.

\bibitem{ref6}
Z.~Zhang et~al., ``Active {RIS} vs. passive {RIS}: Which will prevail in
  {6G}?'' \emph{{IEEE} Trans. Commun.}, vol.~71, no.~3, pp. 1707--1725, Mar.
  2023.

\bibitem{ref9}
L.~Dong, H.-M. Wang, and J.~Bai, ``Active reconfigurable intelligent surface
  aided secure transmission,'' \emph{{IEEE} Trans. Veh. Technol.}, vol.~71,
  no.~2, pp. 2181--2186, Feb. 2022.

\bibitem{zhi2022activeris}
K.~Zhi, C.~Pan, H.~Ren, K.~K. Chai, and M.~Elkashlan, ``Active {RIS} versus
  passive {RIS}: Which is superior with the same power budget?'' \emph{{IEEE}
  Commun. Lett.}, vol.~26, no.~5, pp. 1150--1154, May. 2022.

\bibitem{ma2022active}
Y.~Ma, M.~Li, Y.~Liu, Q.~Wu, and Q.~Liu, ``Active reconfigurable intelligent
  surface for energy efficiency in {MU-MISO} systems,'' \emph{{IEEE} Trans.
  Veh. Technol.}, vol.~72, no.~3, pp. 4103--4107, Mar. 2023.

\bibitem{2023liperformance}
Q.~Li, M.~El-Hajjar, I.~Hemadeh, D.~Jagyasi, A.~Shojaeifard, and L.~Hanzo,
  ``Performance analysis of active {RIS}-aided systems in the face of imperfect
  {CSI} and phase shift noise,'' \emph{{IEEE} Trans. Veh. Technol.}, vol.~72,
  no.~6, pp. 8140 -- 8145, Jun. 2023.

\bibitem{liuxue2022_D2D}
Z.~Peng, X.~Liu, C.~Pan, L.~Li, and J.~Wang, ``Multi-pair {D2D} communications
  aided by an active {RIS} over spatially correlated channels with phase
  noise,'' \emph{{IEEE} Wireless Commun. Lett.}, vol.~11, no.~10, pp.
  2090--2094, Oct. 2022.

\bibitem{2020badiucommunication}
M.-A. Badiu and J.~P. Coon, ``Communication through a large reflecting surface
  with phase errors,'' \emph{{IEEE} Wireless Commun. Lett.}, vol.~9, no.~2, pp.
  184--188, Feb. 2020.

\bibitem{papazafeiropoulos2021intelligent}
A.~Papazafeiropoulos, C.~Pan, P.~Kourtessis, S.~Chatzinotas, and J.~M. Senior,
  ``Intelligent reflecting surface-assisted {MU-MISO} systems with imperfect
  hardware: Channel estimation and beamforming design,'' \emph{{IEEE} Trans.
  Wireless Commun.}, vol.~21, no.~3, pp. 2077--2092, Mar. 2022.

\bibitem{ref27}
J.-F. Bousquet, S.~Magierowski, and G.~G. Messier, ``A 4-{GH}z active scatterer
  in 130-nm {CMOS} for phase sweep amplify-and-forward,'' \emph{IEEE Trans.
  Circuits Syst. I, Reg. Papers}, vol.~59, no.~3, pp. 529--540, Mar. 2012.

\bibitem{ref31}
E.~Bj{\"o}rnson, J.~Hoydis, and L.~Sanguinetti, ``Massive {MIMO} networks:
  Spectral, energy, and hardware efficiency,'' \emph{Found. Trends Signal
  Process.}, vol.~11, no. 3-4, pp. 154--655, Nov. 2017.

\bibitem{2003hassibihow}
B.~Hassibi and B.~M. Hochwald, ``How much training is needed in
  multiple-antenna wireless links?'' \emph{{IEEE} Trans. Inf. Theory}, vol.~49,
  no.~4, pp. 951--963, Apr. 2003.

\bibitem{peng2022performance}
Z.~Peng, X.~Chen, C.~Pan, M.~Elkashlan, and J.~Wang, ``Performance analysis and
  optimization for {RIS}-assisted multi-user massive {MIMO} systems with
  imperfect hardware,'' \emph{{IEEE} Trans. Veh. Technol.}, vol.~71, no.~11,
  pp. 11\,786--11\,802, Nov. 2022.

\bibitem{ref29}
R.~Long, Y.-C. Liang, Y.~Pei, and E.~G. Larsson, ``Active reconfigurable
  intelligent surface-aided wireless communications,'' \emph{{IEEE} Trans.
  Wireless Commun.}, vol.~20, no.~8, pp. 4962--4975, Aug. 2021.

\bibitem{wu2021intelligent}
Q.~Wu, S.~Zhang, B.~Zheng, C.~You, and R.~Zhang, ``Intelligent reflecting
  surface-aided wireless communications: A tutorial,'' \emph{{IEEE} Trans.
  Commun.}, vol.~69, no.~5, pp. 3313--3351, May. 2021.

\bibitem{zheng2021double}
B.~Zheng, C.~You, and R.~Zhang, ``Double-{IRS} assisted multi-user {MIMO}:
  Cooperative passive beamforming design,'' \emph{{IEEE} Trans. Wireless
  Commun.}, vol.~20, no.~7, pp. 4513--4526, Jul. 2021.

\bibitem{ref35}
S.~M. Kay, \emph{Fundamentals of Statistical Signal Processing}.\hskip 1em plus
  0.5em minus 0.4em\relax Upper Saddle River, NJ, USA: Prentice-Hall, 1993.

\end{thebibliography}

\end{document}